\documentclass[a4paper,graphics,natbib,graphicx,psfig,amssymb,ccaption,layout]{article}
\usepackage{lscape, morefloats , graphicx, float}
\newcommand{\affil}[1]{$^{\rm #1}$}
\usepackage[authoryear]{natbib}
\bibpunct{(}{)}{;}{a}{}{,}
\date{} 

\title{\large\bf\flushleft Local stellar kinematics from RAVE data:
IV. Solar neighbourhood age-metallicity relation}
\author{\parbox{\textwidth}{\flushleft
\vspace{-0.5cm}
{\it \c{S}. Duran\affil{A, *}, S. Ak\affil{A}, S. Bilir\affil{A}, S. Karaali\affil{A}, T. Ak\affil{A}, Z.~F. Bostanc\i\affil{B},  B. Co\c skuno\u glu\affil{A}}\\
\vspace{0.4cm}
{\small \affil{A}\,Istanbul University, Faculty of Sciences, Department of Astronomy and Space Sciences, 34119, Istanbul, Turkey}\\
{\small \affil{B}\,Sabanc\i~University, Faculty of Engineering and Natural   Sciences, 34956, Orhanl\i-Tuzla, Istanbul, Turkey}\\
{\small \affil{*}\,Email: duransivan@gmail.com}}}
\begin{document}
\twocolumn[
\begin{changemargin}{.8cm}{.5cm}
\begin{minipage}{.9\textwidth}
\vspace{-1cm}
\maketitle
\small{\bf Abstract:}
We investigated the age-metallicity relation using 
a sample of 5691 F and G type dwarfs from RAVE DR3 by applying several 
constraints: i) We selected stars with surface gravities 
$\log g~($cm~s$^{-2}) \geq3.8$ and effective temperatures in the  
$5310\leq T_{eff}(K)\leq7300$ range and obtained a dwarf sample. ii) We plotted the 
dwarfs in metallicity sub-samples in the $T_{eff}-(J-K_s)_0$ plane to compare 
with the corresponding data of \citet{Gonzalez09} and identified the ones in 
agreement. iii) We fitted the reduced dwarf sample obtained from constraints 
(i) and (ii) to the Padova isochrones and re-identified those which occupy 
the plane defined by isochrones with ages $t\leq13$ Gyr. iv) Finally, we 
omitted dwarfs with total velocity errors larger than 10.63 km~s$^{-1}$. 
We estimated the ages using the Bayesian procedure of \citet{Jorgensen05}. 
The largest age-metallicity slope was found for early F-type dwarfs.
We found steeper slopes when we plotted the data as a function of spectral 
type rather than Galactic population. We noticed a substantial scatter in 
metallicity distribution at all ages. The metal-rich old dwarfs 
turned out to be G-type stars which can be interpreted as they migrated 
from the inner disc or bulge.

\medskip
\medskip
\end{minipage}
\end{changemargin}
]
\small

\section{Introduction}
Age-metallicity relation (AMR) provides information for our 
understanding the formation and evolution of Milky Way. The pioneering studies 
of \citet{Rocha-Pinto00} and \citet{Twarog80a, Twarog80b} show good 
correlations between age and metallicity. However, recent investigations 
indicate that the picture is more complicated. As time progresses the 
interstellar medium becomes enriched in heavy elements. As a result, the newly 
formed stars should have higher metallicity than the older ones. Contrary to 
this expectation, we observe metal rich young and old stars together 
\citep*[e.g.][]{Edvardsson1993, Carraro1998, Chen00, Feltzing01b}.

\citet{Feltzing01b} separated their sample of 5828 dwarf and sub-dwarf stars 
into five sub-samples in terms of effective temperature, i.e. a) 
$3.83<\log T_{eff}(K)$, b) $3.80< \log T_{eff}(K)\leq3.83$, c) 
$3.77<\log T_{eff}(K)\leq3.80$, d) $3.75<\log T_{eff}(K)\leq3.77$, 
and e) $\log T_{eff}(K)\leq3.75$. The hottest two sub-samples show a good 
AMR, but for cooler sub-samples, old metal-rich 
stars contaminate the expected AMR. Since 
\citet{Feltzing01b} derived accurate metallicities from Str\"omgren 
photometry, their results are reliable.

A similar contamination can be observed in three papers of the Geneva-Copenhagen 
Survey (GCS) group. The advantage of this group is that their radial velocity 
data were obtained homogeneously with the photoelectric cross-correlation 
spectrometers CORAVEL \citep*{Baranne79, Mayor85}. In paper I 
\citep{Nordstrom04}, they derived a relation between age and metallicity 
for young dwarf stars but a substantial scatter in metallicity was present 
at all ages. They stated that reliable ages can not be determined for unevolved 
stars and that the relative errors in individual ages exceed 50 per cent. 
Such high scattering blurs the expected AMR. 

The Geneva-Copenhagen survey group improved their calibrations in the 
following papers. In Paper II \citep*{Holmberg07} they included the $V-K$ 
photometry, whereas in Paper III \citep*{Holmberg09} they used the recent 
revision of {\em Hipparcos} parallaxes \citep*{vanleeuwen07}. The most 
striking feature found in these studies is the existence of metal rich old 
stars together with metal-poor old ones, with large scattering. 

AMR for \citet{Soubiran08}'s 891 clump giants 
is different than the one for dwarfs mentioned in the 
preceding paragraphs. Their results are mainly for the thin 
disc. The number of metal-rich stars is small in their sample. They derived 
a vertical metallicity gradient of -0.31 dex kpc$^{-1}$ and found a 
transition in metallicity at $\sim$4-5 Gyr. The metallicity decreases with increasing 
ages up to $\sim$4-5 Gyr, whereas it has a flat distribution at higher ages. 
Additionally, the metallicity scatter is rather large.      

Although the data used for age-metallicity calibrations 
are improved, we are still far from our aim. Establishing
a global relation between age and metallicity is challenging 
with current data. We deduce from the cited studies that such a relation 
can be obtained only with some constraints, such as temperature, 
spectral type, population, luminosity class etc. Even under these 
limitations, the parameters must be derived precisely. Then, we 
should comment whether the solar neighbourhood is as homogeneous 
as it was during its formation or whether its evolution has been 
shaped by mass accretion or spiral waves.  

In this paper, we use a different sample of F and G dwarfs and 
derive the AMR for several sub-samples. The data were 
taken from RAVE DR3 and are given in Section 2. The main 
sequence sample is identified by two constraints, i.e. 
$\log g \geq 3.8~($cm s$^{-2})$ and $5310 \leq T_{eff}~(K) \leq 7300$. 
Parallaxes are not available for stars observed in RAVE survey. 
Hence, the distances of the sample stars were calculated 
by applying the main-sequence colour-luminosity 
relation of \citet{Bilir08} which is valid in the absolute magnitude 
range $0< M_J <6$. We de-reddened the colours and magnitudes by using 
an iterative process and the maps of \citet*{Schlegel98} plus the 
canonical procedure appearing in the literature (see Section 2.2). 
We combined the distances, the RAVE kinematics and the available 
proper motions to estimate the $U,~V,~W$ space velocity components 
which will be used for population analyses. Four types of populations are 
considered, i.e. high probability thin disc stars, low probability 
thin disc stars, low probability thick disc stars, and high probability 
thick disc stars. The separation of the sample stars into different 
population categories is carried out by the procedures in \citet*{Bensby03} 
and \citet{Bensby05}. In Section 3, we used the procedure of \citet*{Jorgensen05} 
which is based on posterior joint probability for accurate 
stellar age estimation. AMR for 32 sub-samples are given in Section 4. 
The sub-samples consist of different spectral and population types 
and their combination. Section 5 is devoted to summary and discussion.

\section{Data}

The data used in this study are from RAVE's third data release 
\citep[DR3;][]{Siebert11}. RAVE DR3 consists of 82~850 stars, each 
with equatorial and Galactic coordinates, radial velocity, metallicity, 
surface temperature and surface gravity. We also note the two former data 
releases, i.e. DR1 \citep{Steinmetz06} and DR2 \citep{Zwitter08}. 
Proper motions were compiled from several catalogues: {\em Tycho-2}, 
Supercosmos Sky Survey, Catalog of Positions and Proper motions on the 
ICRS \citep*[PPMXL;][]{Roeser10} and USNO CCD Astrograph Catalog 2 
(UCAC-2). Proper motion accuracy decreases in this order, therefore, 
if proper motions were available from all catalogues, {\em Tycho-2}'s 
values were used. If {\em Tycho-2} did not provide proper motions, then 
the values were taken from the Supercosmos Sky Survey, etc. Photometric 
data are based on near-IR (NIR) system. The magnitudes of stars were 
obtained by matching RAVE DR3 with Two Micron All Sky Survey 
\citep[2MASS;][]{Skrutskie06} point source catalogue \citep{Cutri03}. 

\subsection{The Sample}
We applied several constraints to obtain a F and G type main sequence 
sample from RAVE DR3. First, we selected stars 
with surface gravities $\log g~($cm~s$^{-2})\geq3.8$ and 
effective temperatures in the $5310\leq T_{eff}(K)\leq7300$ range \citep{Cox00}. 
Thus, the sample was reduced to 18~225 stars. The distribution of all DR3 
stars in the $\log g$, $\log T_{eff}$ plane is given in Fig. 1. 
The F and G main-sequence sample used in this study is also 
marked in the figure. Then, we separated the star sample into 
four metallicity intervals, i.e. $0<[M/H]\leq0.5$, 
$-0.5<[M/H]\leq0$,$-1.5<[M/H]\leq-0.5$,$-2.0<[M/H]\leq-1.5$ dex, 
and plotted them in the $T_{eff}-(J-K_s)_{0}$ plane (Fig. 2) to
compare them with the data of \citet{Gonzalez09}. 

Next, we separated the sample into four metallicity intervals, 
i.e. $0.2\leq [M/H]$, $-0.2\leq[M/H]<0.2$,$-0.6\leq[M/H]<-0.2$, $[M/H]<-0.6$ 
dex, and plotted them in the $\log g$-$\log T_{eff}$ plane in order to 
compare their positions with the zero-age-main sequence (ZAMS) Padova 
isochrone (Fig. 3). The calculation of age by using a set of 
isochrones for stars with masses $0.15<M_{\odot}\leq100$, 
metal abundances $0.0001\leq Z\leq 0.03$ and ages from $\log (t/yr)\leq10.13$ 
is published on the web site of the Padova research 
group\footnote{http://stev.oapd.inaf.it/cgi-bin/cmd} and described in 
the work of \citet{Marigo08}. We omitted 2669 stars which fall below the 
ZAMS, thus the sample was reduced to 8119 stars. Finally, we fitted our 
stars to the Padova isochrones with ages 0, 2, 4, 6, 8, 10, 12, 13 Gyr and 
metallicities $0.2\leq[M/H]$, $0\leq[M/H]<0.2$,$-0.2\leq[M/H]<0$,
$-0.4\leq[M/H]<-0.2$, $-0.6\leq[M/H]<-0.4$, $[M/H]<-0.6$ dex in Fig. 4, 
and excluded the stars not between the isochrones. Thus, we obtained a final 
sample with 6545 F and G main-sequence stars.

Fig. 2 shows that there are substantial differences between the  
$T_{eff} - (J-K_s)_0$ relations for four metallicity intervals of 
our sample and the one of \citet*{Gonzalez09}. 
As the $(J-K_s)_0$ colours are accurate the difference in question 
should  originate from the temperatures, i.e. the temperature 
errors of the sample stars are larger than the mean temperature, 
$\Delta T \sim 200~K$, cited by \citet{Siebert11}. If we decrease the 
temperatures of the sample stars such as to fit with the ones of 
\citet*{Gonzalez09}, their positions move to 
lower temperatures as in Fig. 3 and Fig. 4.

One can notice a large fraction of stars are below the ZAMS in 
Fig. 3. The surface gravities of these stars extend up to 
$\log g = 5.0~($cm s$^{-2})$ which is not expected. If we decrease 
the effective temperatures of the sample stars as mentioned 
in the previous paragraph, the result does not change considerably. 
Hence, the only explanation of the large fraction of stars in 
question can be that the errors in surface gravities are (probably) 
larger than the mean errors cited by \citet{Siebert11}, 
i.e. $\Delta \log g = 0.2~($cm s$^{-2})$. A decrease of 
$\Delta \log g \sim 0.5~($cm s$^{-2})$
in the surface gravity moves the sample stars to an agreeable 
position in Fig. 3. Such a revision provides also a much better 
fitting of the sample stars to the isochrones in Fig. 4.

Now, there is a question to be answered: How does such a revision 
affect the AMR, our main goal in this study? 
Although a decrease in the effective temperature increases the number 
of sample stars (Fig. 4) not on the isochrones, the reduction of their 
surface gravities removes them to an agreeable position in the 
($\log g$, $\log T_{eff}$) plane. Most of the sample stars (if not all) 
are (thin or thick) disc stars. Moreover, age does 
not depend on surface gravity. Hence, any revision in temperature or in 
surface gravity will \textit{increase the number} of the sample stars 
at a given position of the AMR \textit{but not 
its trend}, i.e. we do not expect any considerable difference in 
our results if we apply such a revision.

\subsection{Distance Determination}

Contrary to the {\em Hipparcos} catalogue \citep{ESA97}, 
parallaxes are not available for stars observed in RAVE survey 
\citep{Steinmetz06}. Hence, the distances of the sample stars 
were calculated using another procedure. We applied the main 
sequence colour-luminosity relation of \cite {Bilir08} which 
is valid in the $0<M_J<6$ range, where $M_J$ is the absolute 
magnitude in $J$-band of 2MASS photometry.  The errors of 
the distances were estimated combining the internal errors 
of the coefficients of \citet{Bilir08}'s equation and the 
errors of the 2MASS colour indices. 

As most of the stars in the sample are at distances larger 
than 100 pc, their colours and magnitudes are affected by 
interstellar reddening. Hence, distance determination is 
carried out simultaneously by de-reddening of the sample stars. 
As a first step in an iterative process, we assume the original 
$(J-H)$ and $(H-K_s)$ colour indices are de-reddened, and evaluate the 
$M_J$ absolute magnitude of the sample stars by means of the 
colour-luminosity relation of \citet {Bilir08}. Combination of 
the apparent and absolute magnitudes for the $J$ band gives the 
distance of a star. We used the maps of \citet*{Schlegel98} and 
evaluated the colour excess $E(B-V)$ for each sample star. 
The relation between the total and selective absorptions in 
the UBV system, i.e. 

\begin{equation}
A_{\infty}(b)=3.1\times E_{\infty}(B-V)
\end{equation}
gives $A_{\infty}(b)$ which can be used in evaluating $A_d(b)$ 
using Bahcall \& Soneira's (1980) procedure: 
\begin{equation}
A_{d}(b)=A_{\infty}(b)\times \Biggl[1-\exp\Biggl(\frac{-\mid
d~\sin(b)\mid}{H}\Biggr)\Biggr]
\end{equation} 
where $b$ and $d$ are the Galactic latitude and distance of 
the star, respectively. $H$ is the scaleheight for the interstellar 
dust which is adopted as 125 pc \citep{Marshall06}, and $A_{\infty}(b)$ and
$A_{d}(b)$ are the total absorptions for the model and for the distance 
to the star, respectively. Then, the colour excess at the distance of 
the star, $E_{d}(B-V)$, can be evaluated using a specific form of Eq. (1):
\begin{equation}
A_{d}(b)=3.1\times E_{d}(B-V)
\end{equation} 
The reduced colour excess was used in Fiorucci \& Munari's (2003) 
equations to obtain
the total absorptions for the $J$, $H$ and $K_s$ bands, i.e. $A_J$ =
0.887 $\times$ $E(B-V)$, $A_H$ = 0.565 $\times$ $E(B-V)$ and $A_{Ks}$
= 0.382 $\times$ $E(B-V)$, which were used in Pogson's equation 
($m_{i}-M_{i}=5\log d-5+A_{i}$; $i$ denotes a specific band) to
estimate distances. Contrary to the assumption above, the original
$(J-H)$ and $(H-K_s)$ colour indices are not de-reddened. Hence, the
application of the equations (1) to (3) is iterated until the
distance $d$ and colour index $E_{d}(B-V)$ approach constant values. 

The distribution of the distances (Fig. 5) shows that $\sim$80 per cent 
of the sample stars have almost a normal distribution within the distance 
interval $0<d\leq0.4$ kpc, whereas the overall distribution which extends 
up to 1 kpc is skewed, with a median of 0.25 kpc. However, $\sim$99 per cent of 
the sample stars are within $d=0.6$ kpc. We compared the distances for a 
set of stars estimated in our study with the ones evaluated by their 
parallaxes to test our distances as explained in the 
following. There are 51 main-sequence stars with small relative parallax errors, 
$\sigma_{\pi}/\pi\leq0.20$, common in RAVE DR3 and newly reduced 
{\em Hipparcos} catalogue \citep{vanleeuwen07} for which one can evaluate 
accurate distances. The constraints applied  for providing this sample are 
as follows: $M_V>4$, $\log g>3.80$ (cm~s$^{-2}$) and $\sigma_{\pi}/\pi\leq0.20$. Fig. 6 
shows that there is a good agreement between two sets of distances. The mean 
and the standard deviations of the residuals are 0 and 11 pc, respectively.
The number of stars common in the {\em Hipparcos} catalogue and in RAVE DR3 decreases, 
while their scatter grows with distance becoming 25 per cent of the distance at 
100 pc. The parallaxes in the {\em Hipparcos} catalogue are not available to 
estimate the scatter at larger distances such as $600$ pc, the largest 
distance reported in our study.

The positions of the sample stars in the rectangular coordinate system 
relative to the Sun are given in Fig. 7. The projected shapes both on 
the Galactic ($X$, $Y$) plane and the vertical ($X$, $Z$) plane of 
the sample show asymmetrical distributions. The median coordinates 
($X$=61, $Y$=$-$97, $Z$=$-$114 pc) of the sample stars confirm this 
appearance. The inhomogeneous structure is due to the incomplete 
observations of the RAVE project and that the programme stars were 
selected from the Southern Galactic hemisphere \citep{Steinmetz06}. 

\subsection{Kinematics}
We combined the distances estimated in Section 2.2 with RAVE kinematics
and available proper motions, applying the algorithms and the 
transformation matrices of \cite{Johnson87} to obtain their Galactic 
space velocity components ($U$, $V$, $W$). In the calculations, the epoch 
of J2000 was adopted as described in the International Celestial 
Reference System (ICRS) of the {\em Hipparcos} and {\em Tycho-2} 
Catalogues \citep{ESA97}. The transformation matrices use the notation 
of a right handed system. Hence, $U$, $V$ and $W$ are the components of 
a velocity vector of a star with respect to the Sun, where $U$ is positive
towards the Galactic centre ($l$=0$^{\circ}$, $b$=0$^{\circ}$), $V$ is
positive in the direction of Galactic rotation ($l$=90$^{\circ}$,
$b$=0$^{\circ}$) and $W$ is positive towards the North Galactic Pole
($b$=90$^{\circ}$). 

Correction for differential Galactic rotation is necessary for
accurate determination of $U$, $V$ and $W$ velocity components. The
effect is proportional to the projection of the distance to the stars
onto the Galactic plane, i.e. the $W$ velocity component is not
affected by Galactic differential rotation \citep{Mihalas81}. We
applied the procedure of \cite{Mihalas81} to the distribution of the
sample stars in the $X$-$Y$ plane and estimated the first order
Galactic differential rotation corrections for $U$ and $V$ velocity
components of the sample stars. The range of these corrections is
-25.4$<$$dU$$<$11 and -2.1$<$$dV$$<$1.6 km s$^{-1}$ for $U$ and
$V$, respectively. As expected, $U$ is affected more than the $V$
component. Also, the high values for the $U$ component show that
corrections for differential Galactic rotation can not be ignored.

The uncertainty of the space velocity components $U_{err}$, $V_{err}$
and $W_{err}$ were computed by propagating the uncertainties of the
proper motions, distances and radial velocities, again using an 
algorithm by \cite{Johnson87}. Then, the error for the
total space velocity of a star follows from the equation:

\begin{equation}
S_{err}^{2}=U_{err}^{2}+V_{err}^{2}+W_{err}^{2}. 
\end{equation}
 
The distributions of errors for both the total space velocity and 
its components are plotted in Fig. 8. The median and standard
deviation for space velocity errors are \~ S$_{err}$=3.76 km s$^{-1}$
and $s$=2.29 km s$^{-1}$, respectively. We now remove the most
discrepant data from the analysis, knowing that outliers in a survey
such as this will preferentially include stars which are systematically
mis-analysed binaries, etc. Thus, we omit stars with errors that
deviate by more than the sum of the standard error and three times of the 
standard deviation, i.e. $S_{err}>10.63$ km s$^{-1}$. This removes 854 
stars, 13 per cent of the sample. Thus, our sample was reduced to
5691 stars, those with more robust space velocity components. After
applying this constraint, the median values and the standard
deviations for the velocity components were reduced to (\~U$_{err}$,
\~V$_{err}$, \~W$_{err}$)=(3.05$\pm$1.56, 2.71$\pm$1.40,
2.38$\pm$1.31) km s$^{-1}$. The two dimensional distribution of the
velocity components for the reduced sample is given in Fig. 9.

\subsection{Population Analysis}
We now wish to consider the population kinematics as a function of
stellar population, using space motion as a statistical process to
label stars as members of a stellar population. We
used the procedure of \cite*{Bensby03} and \cite{Bensby05} to allocate
the main sequence sample (5691 stars) into populations and derived
the solar space velocity components for the thin disc population to
check the dependence of Local Standard of Rest (LSR) parameters on 
population. \cite{Bensby03, Bensby05} assumed that the Galactic space 
velocities of stellar populations with respect to the LSR have Gaussian 
distributions as follows:

\begin{equation}
f(U,~V,~W)=k~\times~\exp\Biggl(-\frac{U_{LSR}^{2}}{2\sigma_{U{_{LSR}}}^{2}}-\frac{(V_{LSR}-V_{asym})
^{2}}{2\sigma_{V{_{LSR}}}^{2}}-\frac{W_{LSR}^{2}}{2\sigma_{W{_{LSR}}}^{2}}\Biggr),
\end{equation}
where 
\begin{equation}
k=\frac{1}{(2\pi)^{3/2}\sigma_{U{_{LSR}}}\sigma_{V{_{LSR}}}\sigma_{W{_{LSR}}}}
\end{equation}
normalizes the expression. For consistency with other analyses, we
adopt $\sigma_{U{_{LSR}}}$, $\sigma_{V{_{LSR}}}$ and
$\sigma_{W{_{LSR}}}$ as the characteristic velocity dispersions: 35,
20 and 16 km s$^{-1}$ for thin disc ($D$); 67, 38 and 35 km s$^{-1}$
for thick disc ($TD$); 160, 90 and 90 km s$^{-1}$ for halo ($H$),
respectively \citep{Bensby03}. $V_{asym}$ is the asymmetric drift:
-15, -46 and -220 km s$^{-1}$ for thin disc, thick disc and halo,
respectively. $U_{LSR}$, $V_{LSR}$ and $W_{LSR}$ are LSR
velocities. The space velocity components of the sample stars relative
to the LSR were estimated by adding the values for the space velocity
components to the corresponding solar ones evaluated by \cite{Coskunoglu11}.

The probability of a star of being ``a member'' of a given population is
defined as the ratio of the $f$($U$, $V$, $W$) distribution functions
times the ratio of the local space densities for two
populations. Thus,
 
\begin{equation}
TD/D=\frac{X_{TD}}{X_{D}}\times\frac{f_{TD}}{f_{D}}~~~~~~~~~~TD/H=\frac{X_{TD}}{X_{H}}\times\frac{f_{TD}}{f_{H}}
\end{equation}
are the probabilities for a star of it being classified as a thick
disc star relative to being a thin disc star, and relative to
being a halo star, respectively. $X_{D}$, $X_{TD}$ and $X_{H}$ are
the local space densities for thin disc, thick disc and halo,
i.e. 0.94, 0.06, and 0.0015, respectively \citep{Robin96,Buser99,Bilir06}. We
followed the argument of \cite{Bensby05} and separated the sample
stars into four categories: $TD/D$$\leq$0.1 (high probability thin
disc stars), 0.1$<$$TD/D$$\leq$1 (low probability thin disc stars),
1$<$$TD/D$$\leq$10 (low probability thick disc stars) and $TD/D$$>$10
(high probability thick disc stars). The distribution of number of stars 
and the Galactic space velocity components for different stellar 
population categories are given in Table 1 and Fig. 10. 

\section {Stellar Age Estimation}
We used the procedure of \citet{Jorgensen05} for stellar age estimation. 
We give a small description of the procedure in this study. We quote 
\citet{Jorgensen05}'s paper for details. This procedure is based on the 
-so called- (posterior) joint probability density 
function as defined in the following:
\begin{equation} 
f(\tau,\zeta,m) \propto f_0(\tau,\zeta,m)~ L(\tau,\zeta,m) 
\end{equation}
where $f_0$ is the prior probability density of the parameters and 
$L$ the likelihood function. The parameters $\tau$, $\zeta$, $m$ 
are the age, initial metallicity and initial mass, respectively. 
The probability density function (pdf) is defined such that 
$f(\tau,\zeta,m)\mbox{d}\tau\mbox{d}\zeta\mbox{d}m$ is 
the fraction of stars with ages between $\tau$ and $\tau+\mbox{d}\tau$, 
metallicities between $\zeta$ 
and $\zeta+\mbox{d}\zeta$, and initial masses between $m$ and $m+dm$. 
The constant of proportionality in Eq. (8) 
must be chosen to make 
$\int\!\int\!\int f(\tau,\zeta,m)\mbox{d}\tau\mbox{d}\zeta\mbox{d}m=1$ 

The likelihood function ($L$) equals the probability of getting 
the observed data $q$($\log T_{eff}$, $\log g$, $[M/H]$) for given parameters $p(\tau,\zeta,m)$. Then, 
the likelihood function is 
\begin{equation} 
L(\tau,\zeta,m) = \left( \prod_{i=1}^n \frac{1}{(2\pi)^{1/2}\sigma_i} \right) \times \exp(-\chi^2/2), 
\end{equation}
where
\begin{equation}
\chi^2 = \sum_{i=1}^n \left(\frac{q_i^{\rm obs}-q_i(\tau,\zeta,m)}{\sigma_i}\right)^2,
\end{equation}
and where $\sigma_i$ is the standard error. A maximum-likelihood (ML) 
estimate of the stellar parameters $(\tau,\zeta,m)$ may be obtained by 
finding the maximum of this function, which is equivalent to minimizing 
$\chi^2$ in the case of Gaussian errors \citep{Jorgensen05}. 

The prior density of the model parameters in Eq. (8) can be written as
\begin{equation}
f_0(\tau,\zeta,m) = \psi(\tau)\phi(\zeta\vert\tau)\xi(m\vert\zeta,\tau),
\end{equation}
where $\psi(\tau)$ is the a priori star formation rate (SFR) history, 
$\phi(\zeta\vert\tau)$ the metallicity distribution as a function of age, 
and $\xi(m\vert\zeta,\tau)$ the a priori initial mass function (IMF) 
as a function of metallicity and age. 

Following \citet{Jorgensen05}, we adopted 
the metallicity distribution as a flat function and a power-law 
\begin{equation}
\xi(m) \propto m^{-\alpha},
\end{equation}
for the initial mass function with $\alpha=2.7$. 
If we insert Eq. (11) into Eq. (8) and integrate 
with respect to $m$ and $\xi$, the posterior pdf of $\tau$ can be written as 
\begin{equation}
f(\tau) \propto \psi(\tau)G(\tau),
\end{equation}
where
\begin{equation}
G(\tau) \propto \int\!\int L(\tau,\zeta,m)\xi(m)~\mbox{d}m~\mbox{d}\zeta.
\end{equation}
We normalize Eq. (14) such that $G(\tau)=1$ at its maximum. 
\citet{Jorgensen05} interpreted $G(\tau)$ as the relative 
likelihood of $\tau$ after eliminating $m$ and $\zeta$.

Following \citet{Jorgensen05}, we evaluated Eq. (14) 
for each age value ($\tau_i$) as a double sum along a set of 
isochrones at the required age that are equidistant in metallicity 
($\zeta_k$). In practice, we used pre-computed isochrones for a 
step size of 0.05 dex in $\zeta$, and considered only those within 
$\pm3.5\sigma_{[M/H]}$ of the observed metallicity. Let $m_{jkl}$ be 
the initial-mass values along each isochrone ($\tau_j$, $\zeta_k$); then 
\begin{equation}
G(\tau_j) \propto \sum_k \sum_\ell L(\tau_j,\zeta_k,m_{jk\ell}) \xi(m_{jk\ell})(m_{jk\ell+1}-m_{jk\ell-1}).
\end{equation}
Age corresponding to the mode of the relative posterior 
probability $G(\tau)$ is adopted as the age of the star in question. 
The distribution of the ages for the final sample (5691 stars) is 
given in Fig. 11.

We tested the ages estimated in this study by comparing them with the ones 
estimated in the GCS by using the procedure explained in the following. RAVE and 
GCS surveys have 142 stars in common. The $T_{eff}$, $M_V$, and $[M/H]$ 
parameters of 66 stars in this sample were determined by \citet{Holmberg09}. 
As we considered only the main-sequence stars, we applied the constraint $M_V>4$. 
Thus the sample reduced to 25 stars. We estimated the ages of these stars by using
PARAM\footnote{http://stev.oapd.inaf.it/cgi-bin/param} webpage and compared them 
with the ages estimated by the procedure in our study. The results are given in 
Fig. 12. Although the final sample consists of a limited number of stars, there is 
a good agreement between the two sets of ages. The errors for all the stars in our 
study are fitted to a Gaussian distribution in Fig. 13. The mode and the 
standard deviation of the distribution are 0.81 and 1.25 Gyr, respectively. 
The ages as well as other stellar parameters of the final sample can be 
provided from Table 2, which is given electronically.

\section{Age-metallicity relation}
In this section, we investigate the age-metallicity relation (AMR). 
The distribution of metallicities for our sample is given in Fig. 14 
as a function of spectral type. The distributions for F and G 
spectral types and their combination give the indication of 
a Gaussian distribution with slightly different modes, i.e. 
$\sim$-0.31, $\sim$-0.20, and $\sim$-0.29 dex for F and G types, 
and their combination. F stars are intrinsically brighter than G stars, 
so sample larger distances and thus include a higher proportion of 
thick disc stars, which shifts the $[M/H]$ distribution to 
metal-poor regions. The normalized metallicity distribution for 
both spectral types shows an expected slight shift of the mode 
as a function of population (Fig. 15). This shows that RAVE DR3 
metallicities are approximately correct. However, Fig. 26 in 
\citet{Nordstrom04} shows that the RAVE DR3 distribution appears to be missing 
a metal-poor tail that should make the metallicity distribution more 
asymmetric. RAVE DR3 $[M/H]$ metallicities have an improved calibration 
compared to RAVE DR2, but this shows that it still needs to be calibrated 
robustly to a $[Fe/H]$ metallicity scale. Given these metal-poor stars are 
a minority, a minority of individual stellar ages derived from these 
incorrectly derived metallicities will also be incorrect. Nevertheless,  
the majority of metallicities and ages will be approximately correct, 
which is adequate for our goal of investigating general statistical trends 
in the AMR.

The metallicity distribution of all sample stars in terms of 
age in Fig. 16 reminds us the complicated picture claimed in 
Section 1: there are metal-rich old stars in additional to 
the young ones. Fig. 16 is qualitatively similar to \citet{Nordstrom04}'s 
Fig. 27. We separated our sample into six sub-samples, 
i.e. F0-F3, F3-F6, F6-F9, F9-G2, G2-G5, G5-G8, and investigated 
the AMR for each sub-sample (Fig. 17). The most conspicuous feature 
in Fig. 17 is the absence of old metal-rich stars in the relatively 
early spectral types. These stars appear in the F9-G2 spectral types 
and dominate later spectral types. The second feature in the distributions 
in Fig. 17 is the slope which decreases (absolutely) gradually when one 
goes from sub-sample F0-F3 to the sub-samples including stars from 
later spectral types and becomes almost zero for the sub-sample G5-G8. 
Finally, the third feature in Fig. 17 is the different range of the age, 
i.e. $0<t\leq3$, $t\leq6$, and $t\leq13$ Gyr, for F0-F3, F3-F6, and F6-G8 
spectral types, respectively. Additionally, the number of old stars 
increases for later spectra types, as expected. 

We investigated the relation between age and metallicity also as 
a function of population, i.e. $TD/D\leq0.1$, $0.1<TD/D\leq1$, $1<TD/D\leq10$, 
and $10<TD/D$. The distributions are given in Fig. 19 for F and G types 
individually. One can notice similar metallicity distributions for F and G 
type stars. Small slopes are apparent, with the exception of F type
stars with $TD/D>10$, i.e. high-probability thick disc stars, for which the 
distribution is flat. As in Fig. 17, the majority of the old, metal-rich 
stars are of G spectral type. 

We applied both constraints stated above and plotted the metallicity of the 
sample stars versus their age. Thus, AMR for F0-F3, F3-F6, F6-F9, F9-G2, 
G2-G5, and G5-G8 sub-samples are now plotted for the population types 
$TD/D\leq0.1$, $0.1<TD/D\leq1$, $1<TD/D\leq10$ (Fig. 20). Comparison of the 
distributions in Figs. 19 and 20 shows that population is not a strong 
indicator for an AMR, whereas, metallicity-age distribution for a series 
of narrow spectral type intervals show a slope.

In Section 2, we mentioned the errors of the effective temperatures and 
the surface gravities of the stars. We investigated their effects on our 
final results, i.e. the AMR, by using the procedure 
explained in the following. We used the metallicity of a sample star and 
estimated its surface gravity, effective temperature, and age simultaneously 
by means of the Padova isochrones. We called it ``artificial age''. A second age 
has been determined using  the metallicity, surface gravity, and effective 
temperature in question plus the corresponding mean errors stated for the 
RAVE DR3 survey. The second set of ages, called ``RAVE ages'', were estimated by 
the procedure of \citet{Jorgensen05} as explained in Section 3. 
We obtained AMR for two sets of ages for 32 sub-samples 
and evaluated the differences between two ages (the residuals) for a given 
metallicity for each sub-sample. We plotted only six of the AMR 
for the ``{\em artificial ages}'', i.e. those for the spectral types 
F0-F3, F3-F6, F6-F9, F9-G2, G2-G5, and G5-G8, in Fig. 18 just as an 
example, but we evaluated the mean and  standard deviations for the residuals 
of all sub-samples. Table 3 shows that the mean residuals and the 
corresponding standard deviations are smaller for the F type stars 
than the G type ones. 

\citet{Jorgensen05} stated that the errors of the stellar ages 
in their study are at least 25 per cent of the corresponding age. A similar 
result can be found in \citep{Nordstrom04}, i.e. the stellar age errors 
of 9428 stars in their study are less than 50 per cent of their ages while 
those for 5472 stars are less than 25 per cent. We evaluated 25 per cent 
of the mean age for each sub-sample in Table 3 (last column) and compared 
them with the (absolute) sum of the corresponding mean age residuals, 
$<\Delta Age>$ and standard deviations, $\sigma$. The sum of two statistics are less 
than 25 per cent of the mean age for all sub-samples, except the sub-sample 
G5-G8. Thus, we can say that the estimated ages and the AMR are confident.

\section{Summary and Discussion}
We applied the following constraints to the RAVE DR3 data consisting of 82~850 
stars and obtained the AMR for 5691 F and G spectral 
type stars: i) We selected stars with surface gravities 
$\log g~($cm~s$^{-2})\geq3.8$ and effective 
temperatures $5310\leq T_{eff} (K)\leq7300$. These are the surface 
gravity range of the main-sequence stars and the temperature range 
of F and G type stars, respectively. ii) We separated the stars into 
the metallicity intervals $0<[M/H]\leq0.5$, $-0.5<[M/H]\leq0$, 
$-1.5<[M/H]\leq-0.5$, $-2<[M/H]\leq-1.5$ dex, and plotted them 
in the $T_{eff}$-$(J-K_s)_0$ plane compared to the data of 
\citet{Gonzalez09}. Then, we omitted stars which did not fit 
the temperature-colour plane of \citet{Gonzalez09}. iii) We separated 
the remaining stars into $0.2\leq[M/H]$, $-0.2\leq[M/H]<0.2$, 
$-0.6\leq[M/H]<-0.2$, $[M/H]<-0.6$ dex metallicity intervals and plotted 
them in the $\log g$-$\log T_{eff}$ plane in order to compare their positions 
with the ZAMS of Padova isochrones. We omitted the stars which 
fell below the ZAMS. iv) We fitted the remaining stars to the Padova ischrones 
with ages 0, 2, 4, 6, 8, 10, 12, 13 Gyr and metallicities $0.2\leq[M/H]$, 
$0\leq [M/H]<0.2$, $-0.2\leq[M/H]<0$, $-0.4\leq[M/H]<-0.2$,
$-0.6\leq[M/H]<-0.4$, $[M/H]<-0.6$ dex, and excluded the stars with 
positions beyond the $\log g$-$\log T_{eff}$ plane occupied by the 
isochrones from the sample. v) Finally, we omitted the stars with 
total velocity error $S_{err}>10.63$ km s$^{-1}$. After these constraints 
the sample was reduced to 5691 F and G type main sequence stars. 

The distances of the sample stars were determined by the colour-luminosity 
relation of \citet{Bilir08}, and the $J$, $H$, and $K_s$ magnitudes were 
de-reddened by the procedure in situ and the equations of \citet{Fiorucci03}. 
We combined the distances with RAVE kinematics and available proper 
motions, applying the algorithms and the transformation matrices 
of \citet{Johnson87} to obtain the Galactic space velocity components 
($U$, $V$, $W$). We used the procedure of \citet{Bensby03, Bensby05} to 
divide the main-sequence sample (5691 stars) into populations and derived 
the solar space velocity components for the thin and thick discs and halo 
populations to check the AMR on population. We used the 
Bayesian procedure of \citet{Jorgensen05} to estimate stellar ages. This 
procedure is based on the joint probability density function which 
consists of a prior probability density of the parameters and a likelihood 
function, and which claims stellar ages are at least as accurate as those obtained 
with conventional isochrone fitting methods.          

The distribution of metallicities for the whole star sample in terms of age 
gives a complicated picture, as claimed in the literature cited in 
Section 1. The most conspicuous feature is the existence of metal rich old 
stars. Although there is a concentration of stars in the plane occupied by 
(relatively) young stars, one can observe stars at every age and metallicity 
in the age-metallicity plane, whereas, we observe an AMR 
for sub-samples defined by spectral type, i.e. F0-F3, F3-F6, F6-F9, F9-G2, 
G2-G5, G5-G8. However, the slope is not constant for all sub-samples. 
The largest slope belongs to the stars in the 
sub-sample F0-F3. It decreases towards later spectral types and the 
distribution becomes almost flat at G type stars. The ages of stars in 
the F0-F3 and F3-F6 sub-samples are less than 6 Gyr. These sub-samples are 
almost equivalent to the sub-samples defined by the effective temperatures 
$3.83<\log T_{eff} (K)$ and $3.8<\log T_{eff}(K)\leq 3.83$ in Fig. 13 of 
\citet{Feltzing01b}. However, the ages of a few dozen of stars with 
$3.8<\log T_{eff}(K)\leq 3.83$ in their study extends up to 9 Gyr. 
We think that this difference between two studies confirm the benefits
of the procedure used for age estimation in our work. 

Old metal rich stars in our study are of G spectral types. They 
have been investigated in many studies. \citet{Pompeia02} 
identified 35 nearby stars with metallicities $-0.8 \leq [M/H] \leq + 0.4$ dex and 
age $10$-$11$ Gyr, and called them ``bulge-like dwarfs''. \citet{Castro97} 
investigated the $\alpha$- and s-element abundances of nine super metal-rich stars 
in detail where five of them had $[M/H] \geq +0.4$ dex. However, they were unable 
to convincingly assign those stars to a known Galactic population. In our study, 
old metal rich stars have different velocity {\em dispersions} than the 
early-type stars, i.e. F0-F6 spectral type. Although the ranges 
of the space velocity components for two sub-samples are almost the 
same (Fig. 21), F0-F6 spectral type stars are concentrated relatively 
to lower space velocities resulting with smaller velocity dispersions, 
while old metal rich G5-G8 spectral type stars are scattered to larger 
space velocities and therefore have relatively larger space velocity dispersions. 
Numerical values are given in Table 4. Comparison of the space velocity 
dispersions of two sub-samples indicate that {\em old metal-rich} G5-G8 
spectral type stars are members of the thick disc. \citet{Wilson11} favour
the procedure ``gas-rich merger'' for the formation of the thick disc. 
However, they state that a fraction of the thick disc stars could be 
formed by the ``radial migration process''. The old metal-rich sub-sample 
in our study may be the candidates of the second procedure, i.e. they 
could have migrated radially from the inner disc of the Galactic bulge. 
Contrary to the F0-F6 spectral type stars, the G type ones are old enough 
for such a radial migration.

The metallicity distributions with respect to age for different populations, 
i.e. $TD/D\leq0.1$, $0.1<TD/D \leq1$, $1<TD/D\leq10$, and $10<TD/D$ also show 
slopes. However, they are different than the ones for sub-samples 
defined by spectral types. They are smaller compared to the former. 
The application of two constraints, i.e. spectral type and population, 
to our sample results in metallicity distributions with respect to age similar 
to the ones obtained for spectral sub-samples alone which indicates that 
spectral type is more effective in establishing an AMR relative to 
population of a star, i.e. population type is not a strong indicator 
in the AMR.

{\bf Conclusion:} We obtained AMR with the RAVE data which have 
common features as well as differences with {\em Hipparcos}. 
Some differences in features can be explained with different constraints: 1) 
The AMR  for the sub-samples F0-F3 and F3-F6 with stars younger than 6 Gyr 
in our study  has almost the same trend of the AMR for \citet{Feltzing01b}'s F-type stars 
younger than 4 Gyr with temperatures $3.83<\log T_{eff}(K)\leq 3.85$ and 
$3.80<\log T_{eff}(K)\leq 3.83$, respectively. 
They give the indication of a large slope. However, a few dozens of 
stars with age $4<t<9.5$ Gyr and temperature $3.80<\log T_{eff}(K)\leq 3.83$ in 
\citet{Feltzing01b} which have a flat distribution do not appear in our study, 
confirming the accuracy of the ages estimated via Bayesian procedure. 
The trend of the AMR becomes gradually flat when one goes to later spectral 
types in our study or to cooler stars in \citet{Feltzing01b}. The AMR for 
the thin-disc red clump (RC) giants in \citet{Soubiran08} confirms the large 
slope for the young stars and the flat distribution for the older ones, 
$t>5$ Gyr. 2) Substantial scatter in metallicity in our study were observed 
in all related studies \citep[cf.][]{Feltzing01b, Nordstrom04, Soubiran08}. 
3) In our study we revealed that the metal-rich old stars are of G-spectral 
type. These stars could have migrated radially from the inner disc or Galactic 
bulge.

\section{Acknowledgments}
This work has been supported in part by the Scientific 
and Technological Research Council (T\"UB\.ITAK) 210T162 
and Scientific Research Projects Coordination Unit of Istanbul 
University. Project number 14474.

We would like to thank to Dr. Bjarne Rosenkilde J{\o}rgensen, Mr. Tolga Din\c{c}er, and 
Mr. Alberto Lombardo for helping us to improve the code related age estimation.
Funding for RAVE has been provided by: the Australian Astronomical Observatory; 
the Leibniz-Institut fuer Astrophysik Potsdam (AIP); the Australian National 
University; the Australian Research Council; the French National Research 
Agency; the German Research Foundation; the European Research Council 
(ERC-StG 240271 Galactica); the Istituto Nazionale di Astrofisica at Padova; 
The Johns Hopkins University; the National Science Foundation of the USA 
(AST-0908326); the W. M. Keck foundation; the Macquarie University; the 
Netherlands Research School for Astronomy; the Natural Sciences and 
Engineering Research Council of Canada; the Slovenian Research Agency; the 
Swiss National Science Foundation; the Science \& Technology Facilities 
Council of the UK; Opticon; Strasbourg Observatory; and the Universities of 
Groningen, Heidelberg and Sydney. The RAVE web site is at 
http://www.rave-survey.org 

This publication makes use of data products from the Two Micron All
Sky Survey, which is a joint project of the University of
Massachusetts and the Infrared Processing and Analysis
Center/California Institute of Technology, funded by the National
Aeronautics and Space Administration and the National Science
Foundation.  This research has made use of the SIMBAD, NASA's
Astrophysics Data System Bibliographic Services and the NASA/IPAC
ExtraGalactic Database (NED) which is operated by the Jet Propulsion
Laboratory, California Institute of Technology, under contract with
the National Aeronautics and Space Administration.

\begin{table*}
\center 
\caption{Distribution of the sample stars for different stellar population categories.} 
\begin{tabular}{cccccc}
\hline
Sp. Type &  $TD/D\leq0.1$ & $0.1<TD/D\leq1$ & $1<TD/D\leq10$ & $TD/D>10$ & Total \\
\hline
         F &      3733 &        246 &         52 &         19 &      4050 \\
         G &      1430 &        161 &         40 &         10 &      1641 \\
     Total &      5163 &        407 &         92 &         29 &      5691 \\
\hline
        \% &      90.72 &       7.15 &       1.62 &       0.51 &       100 \\
\hline
\end{tabular}
\end{table*} 

\begin{landscape} 
\textwidth = 650pt
\begin{table*}
\setlength{\tabcolsep}{3pt}
{\scriptsize
\caption{Stellar atmospheric parameters, astrometric, kinematic and age data for the whole sample: (1): Our catalogue number, (2): RAVEID, (3-4): Equatorial coordinates in degrees (J2000), (5): $T_{\textrm{eff}}$ in K, (6) $\log g$ (cm~s$^{-2}$) in dex, (7): Calibrated metallicity $[M/H]$ (dex), (8-9): $d$ distance and its error (pc), (10-11) total proper motion and its error in (mas yr$^{-1}$), (12-13): Heliocentric radial velocity and its error in km s$^{-1}$, (14-19): Galactic space velocity components, and their respective errors in km s$^{-1}$, (20): $TD/D$ ratio as mentioned in the text, (20-22): Age and its lower and upper confidential levels.}
\begin{tabular}{ccccccccccccccccccccccccc}
\hline
(1) &  (2) &  (3)  &    (4) & (5) &  (6) & (7) & (8) & (9) & (10) &  (11) & (12) &  (13) &  (14) & (15) & (16) & (17) & (18) & (19) & (20) & (21) & (22) & (23)\\
ID & Designation &  $\alpha$ & $\delta$ & $T_{eff}$ &  $\log g$ & $[M/H]$ & d & d$_{err}$ & $\mu$ &  $\mu_{err}$ & $\gamma$ &  $\gamma_{err}$ & $U$ & $U_{err}$  & $V$ & $V_{err}$ & $W$ & $W_{err}$ & $TD/D$ &  Age & Age$_{lower}$ & Age$_{upper}$\\
\hline
1&	J000014.2-415524&	0.058958&	-41.923389&	6256&	4.27&	-0.38&	148&	16&	6.6&	1.3&	11.2&	1.0&	2.67&	0.73&	2.87&	0.78&	-11.56&	0.98&	0.016&	5.83&	5.11&	6.52&	\\
2&	J000018.6-094054&	0.077375&	-9.681528&	6442&	4.31&	-0.04&	345&	37&	7.7&	2.9&	-5.5&	1.3&	4.72&	3.40&	-10.88&	3.31&	1.91&	1.72&	0.005&	1.47&	1.47&	4.02&	\\
3&	J000025.3-471047&	0.105333&	-47.179583&	6406&	3.94&	-0.64&	218&	24&	12.6&	2.4&	0.5&	1.1&	6.30&	1.77&	11.74&	2.10&	-1.55&	1.23&	0.026&	5.60&	5.11&	6.59&	\\
4&	J000026.3-394149&	0.109500&	-39.696861&	6713&	4.16&	-0.39&	259&	28&	24.3&	2.0&	24.9&	0.8&	35.94&	3.26&	-3.89&	1.72&	-15.79&	1.20&	0.028&	2.71&	2.49&	2.89&	\\
5&	J000033.7-483343&	0.140542&	-48.561889&	6558&	4.25&	-0.32&	345&	37&	20.3&	3.0&	23.1&	1.8&	39.95&	4.43&	5.31&	3.65&	-13.27&	2.30&	0.053&	3.57&	3.23&	3.97&	\\
6&	J000034.3-372148&	0.143042&	-37.363250&	6003&	4.15&	-0.47&	121&	14&	18.2&	1.5&	-9.4&	0.9&	7.83&	1.22&	2.30&	0.62&	11.54&	0.96&	0.013&	11.68&	10.57&	12.78&	\\
7&	J000039.7-484253&	0.165250&	-48.714639&	6679&	3.92&	-0.15&	263&	29&	19.7&	2.8&	20.2&	1.3&	-13.65&	3.23&	0.93&	2.51&	-27.85&	1.79&	0.046&	1.93&	1.68&	2.10&	\\
8&	J000042.6-495124&	0.177333&	-49.856556&	5918&	4.37&	-0.25&	236&	26&	59.9&	2.8&	1.1&	1.9&	-20.93&	3.15&	-62.81&	6.54&	7.99&	2.10&	0.012&	8.41&	6.19&	10.07&	\\
9&	J000129.9-595545&	0.374667&	-59.929194&	6454&	4.47&	-0.53&	254&	27&	25.6&	4.2&	-31.0&	0.8&	-37.31&	4.29&	11.27&	3.33&	13.01&	2.45&	0.036&	0.16&	0.00&	2.69&	\\
10&	J000148.7-075808&	0.452875&	-7.968944&	6514&	3.96&	-0.38&	315&	34&	18.9&	2.7&	14.4&	1.2&	22.84&	3.75&	-4.52&	2.90&	-17.25&	1.62&	0.020&	2.71&	2.42&	3.14&	\\
11&	J000204.1-483954&	0.516875&	-48.665000&	6359&	4.29&	-0.02&	589&	64&	16.9&	3.3&	37.0&	1.7&	56.15&	7.28&	-25.36&	6.45&	-15.57&	3.49&	0.032&	1.86&	0.42&	2.30&	\\
12&	J000205.3-420859&	0.522250&	-42.149833&	5709&	3.91&	-0.62&	181&	19&	45.1&	2.7&	-2.2&	2.7&	22.16&	2.72&	-29.74&	3.29&	12.92&	2.77&	0.005&	11.84&	9.31&	12.56&	\\
13&	J000213.4-474314&	0.555833&	-47.720500&	6261&	4.20&	-0.42&	386&	42&	25.5&	3.0&	8.1&	1.6&	-39.08&	5.70&	-11.42&	4.01&	-20.29&	2.46&	0.017&	5.52&	4.94&	6.19&	\\
14&	J000214.2-495138&	0.558958&	-49.860528&	5991&	4.28&	-0.23&	433&	47&	14.4&	4.5&	8.1&	2.1&	-21.21&	6.77&	-12.62&	6.45&	-14.22&	3.40&	0.008&	8.33&	7.50&	9.38&	\\
15&	J000215.3-374547&	0.563542&	-37.763028&	6069&	4.33&	-0.20&	358&	39&	6.7&	2.8&	13.6&	1.0&	7.79&	2.84&	-11.83&	3.95&	-11.44&	1.32&	0.007&	6.54&	5.33&	7.29&	\\
16&	J000228.7-405839&	0.619417&	-40.977444&	6293&	4.35&	-0.33&	187&	21&	85.5&	2.5&	-38.3&	0.9&	-74.99&	6.78&	-31.43&	3.96&	22.73&	1.69&	0.035&	4.82&	2.98&	5.43&	\\
17&	J000230.7-531342&	0.627917&	-53.228444&	6539&	4.23&	-0.55&	345&	37&	26.0&	3.5&	1.2&	1.4&	-33.50&	5.24&	-20.61&	4.26&	-8.63&	2.35&	0.006&	4.98&	4.45&	5.49&	\\
18&	J000233.6-531053&	0.639917&	-53.181444&	6247&	4.42&	-0.12&	294&	31&	38.6&	3.3&	11.2&	1.4&	-24.25&	3.96&	-47.88&	5.41&	-6.69&	2.08&	0.006&	0.02&	0.00&	1.79&	\\
19&	J000246.2-673718&	0.692667&	-67.621611&	5752&	4.32&	0.07&	185&	19&	48.8&	2.5&	1.1&	1.1&	-22.39&	2.80&	-34.18&	3.45&	7.40&	1.55&	0.004&	7.86&	6.06&	9.18&	\\
20&	J000257.8-485403&	0.741000&	-48.900806&	6401&	4.48&	-0.36&	184&	20&	42.2&	2.7&	17.3&	0.8&	-22.91&	3.32&	-25.54&	2.64&	-20.69&	1.07&	0.009&	0.00&	0.00&	0.01&	\\
21&	J000319.9-660915&	0.832833&	-66.154083&	5447&	4.50&	-0.02&	186&	20&	97.0&	3.8&	12.2&	1.0&	-67.66&	7.63&	-6.00&	2.13&	-50.09&	4.30&	2.489&	6.38&	6.22&	18.31&	\\
22&	J000323.3-495101&	0.847000&	-49.850222&	5614&	4.48&	0.11&	269&	30&	32.3&	2.0&	5.6&	1.8&	-31.98&	4.00&	-20.91&	2.63&	-12.40&	1.94&	0.007&	0.00&	0.00&	0.01&	\\
23&	J000340.5-410545&	0.918708&	-41.095889&	6974&	4.00&	-0.08&	237&	26&	19.7&	1.3&	-0.1&	2.8&	-9.12&	1.56&	-19.94&	2.22&	-0.14&	2.65&	0.004&	1.54&	1.46&	1.65&	\\
24&	J000342.7-572902&	0.927833&	-57.483750&	5879&	4.33&	0.03&	379&	40&	21.7&	5.2&	12.0&	1.3&	27.71&	6.46&	-31.84&	6.72&	9.99&	4.12&	0.006&	5.52&	3.52&	6.57&	\\
25&	J000344.9-473243&	0.937000&	-47.545250&	5928&	3.91&	-0.28&	192&	21&	41.5&	2.9&	-0.4&	0.8&	-31.53&	3.66&	-18.03&	2.56&	-6.27&	1.24&	0.005&	6.69&	5.43&	8.43&	\\
26&	J000357.9-200144&	0.991125&	-20.028972&	5778&	4.38&	-0.29&	214&	23&	23.7&	4.2&	-18.1&	1.3&	11.27&	3.26&	-22.91&	3.51&	15.08&	1.45&	0.005&	12.00&	8.73&	14.47&	\\
27&	J000358.2-711643&	0.992292&	-71.278694&	5943&	4.00&	-0.17&	113&	12&	32.2&	1.7&	11.7&	0.7&	6.21&	0.70&	-20.38&	1.48&	2.01&	1.19&	0.004&	7.86&	7.44&	8.73&	\\
28&	J000406.2-762751&	1.025792&	-76.464194&	6229&	4.25&	-0.31&	227&	25&	39.2&	3.0&	2.0&	1.7&	-32.88&	4.35&	-16.02&	2.55&	-13.81&	2.41&	0.008&	5.99&	5.40&	6.58&	\\
29&	J000412.3-505501&	1.051250&	-50.917000&	6138&	4.38&	-0.29&	309&	34&	16.5&	4.5&	38.9&	1.5&	-7.39&	4.98&	-16.55&	4.56&	-41.65&	2.53&	0.102&	5.29&	1.65&	6.16&	\\
30&	J000416.0-443231&	1.066542&	-44.541889&	6310&	4.05&	-0.41&	247&	26&	3.6&	2.9&	-9.6&	0.7&	0.63&	2.28&	-1.71&	2.42&	10.38&	1.08&	0.009&	5.60&	4.99&	6.29&	\\
31&	J000429.2-674002&	1.121583&	-67.667111&	6115&	4.12&	-0.49&	146&	16&	61.0&	2.1&	12.2&	0.7&	-28.72&	3.67&	-27.93&	2.34&	-14.65&	1.03&	0.006&	10.20&	9.25&	11.16&	\\
32&	J000432.4-464515&	1.135000&	-46.754111&	5626&	4.31&	0.15&	288&	31&	27.4&	3.1&	30.3&	1.2&	-18.14&	4.03&	-29.87&	3.71&	-32.29&	1.62&	0.026&	8.72&	6.93&	10.06&	\\
33&	J000440.8-420221&	1.169958&	-42.039222&	6272&	4.22&	-0.59&	118&	13&	61.2&	2.8&	23.8&	0.8&	-26.10&	3.44&	-4.17&	1.10&	-31.90&	1.23&	0.057&	8.56&	7.71&	9.37&	\\
34&	J000451.1-532059&	1.212750&	-53.349639&	6281&	4.27&	-0.51&	145&	16&	88.7&	2.8&	-0.4&	1.4&	-55.88&	5.83&	-12.87&	1.88&	-17.22&	2.26&	0.022&	7.47&	6.60&	8.62&	\\
35&	J000513.0-204557&	1.304167&	-20.765861&	6121&	4.04&	-0.62&	254&	27&	46.2&	3.5&	-27.5&	1.0&	47.55&	5.75&	-25.83&	3.54&	28.23&	1.13&	0.031&	9.58&	8.48&	10.69&	\\
36&	J000529.9-203620&	1.374375&	-20.605611&	5656&	4.43&	-0.03&	312&	33&	28.8&	3.8&	-4.5&	2.9&	-31.69&	4.89&	-30.35&	4.85&	-4.38&	3.03&	0.004&	7.00&	0.93&	11.55&	\\
37&	J000537.2-522806&	1.405167&	-52.468333&	6238&	3.97&	-0.12&	318&	34&	23.4&	3.5&	1.9&	1.5&	16.41&	4.08&	-30.04&	4.35&	13.04&	2.47&	0.005&	2.87&	2.54&	3.22&	\\
38&	J000540.6-663024&	1.419208&	-66.506583&	5913&	4.25&	-0.33&	97&	10&	51.7&	2.1&	-25.9&	0.7&	-30.07&	2.04&	1.64&	1.24&	15.73&	0.80&	0.015&	11.14&	9.71&	12.22&	\\
39&	J000551.8-411637&	1.465750&	-41.276889&	5953&	4.05&	-0.30&	299&	33&	5.2&	2.8&	42.6&	1.4&	18.10&	2.82&	-9.47&	2.84&	-38.60&	1.61&	0.117&	9.03&	7.22&	9.57&	\\
40&	J000600.7-472917&	1.503000&	-47.488083&	6007&	4.45&	-0.22&	309&	33&	10.0&	3.1&	33.6&	1.2&	21.65&	3.23&	3.31&	3.33&	-30.63&	1.66&	0.104&	2.17&	1.17&	6.61&	\\
...&	...&	...&	...&	...&	...&	...&	...&	...&	...&	...&	...&	...&	...&	...&	...&	...&	...&	...&	...&	...&	...&	...&	\\
5689&	J235930.2-342739&	359.875833&	-34.460944&	6389&	3.93&	-0.28&	203&	22&	4.2&	2.4&	18.8&	1.0&	7.74&	1.81&	1.84&	1.43&	-17.57&	1.06&	0.023&	2.79&	2.50&	3.29&	\\
5690&	J235931.8-405629&	359.882417&	-40.941417&	6526&	4.04&	-0.33&	204&	22&	20.0&	2.8&	-3.0&	0.5&	-17.55&	2.52&	-7.35&	2.06&	-1.28&	0.86&	0.006&	3.03&	2.51&	3.68&	\\
5691&	J235952.7-382813&	359.969500&	-38.470167&	6206&	4.27&	-0.08&	308&	33&	22.4&	2.4&	12.5&	1.7&	-25.90&	3.77&	-12.76&	2.72&	-19.04&	1.90&	0.011&	4.43&	3.87&	4.98&	\\
 \hline
\end{tabular}  

}  
\end{table*} 
\end{landscape}  

\begin{table*}
\center 
\caption{Mean and  standard deviations  for the differences 
between the ages estimated by means of two different sets of 
data for different spectral type intervals and for different 
population types (statistics for the combination  of the 
spectral types and population types, not given in this table, 
are not different than the ones for corresponding spectral types). 
The last column gives 25 per cent of the mean age for comparison 
the errors in this study with the ones in the literature (see the text).} 
\begin{tabular}{lccc}
\hline
Sub-sample &  $< \Delta$Age$>$ (Gyr) & $\sigma$ (Gyr) & $0.25 \times t$ (Gyr) \\
\hline
F0-F3	&	-0.1		&	0.3		&	0.38 \\
F3-F6	&	-0.2		&	0.5		&	0.62 \\
F6-F9	&	-0.3		&	0.7		&	1.38 \\
F9-G2	&	-0.1		&	0.4		&	1.62 \\
G2-G5	&	-0.3		&	0.8		&	1.62 \\
G5-G8	&	-0.7		&	1.4		&	1.62 \\
F type, $TD/D \leq 0.1$	&	-0.2		&	0.6	&	1.62 \\
F type, $TD/D \leq 1.0$	&	-0.1		&	0.6	&	1.62 \\
F type, $1 < TD/D \leq 10$	&	-0.2		&	0.4	&	1.62 \\
F type, $10 < TD/D$	& 	-0.2		&	0.3	&	1.62 \\
G type, $TD/D \leq 0.1$	&	-0.4		&	1.0	&	1.62 \\
G type, $TD/D \leq 1.0$	&	-0.4		&	1.0	&	1.62 \\
G type, $1 < TD/D \leq 10$&	-0.2		&	0.5	&	1.62 \\
G type, $10 < TD/D$	& 	-0.3		&	0.5	&	1.62 \\
\hline
\end{tabular}
\end{table*} 

\begin{table*}
\center 
\caption{Space velocity dispersions for two  sub-samples (units in km s$^{-1}$).}
\begin{tabular}{cccccc}
\hline
Sub-sample &  N & $\sigma_U$ & $\sigma_V$ & $\sigma_W$ & $\sigma_T$ \\
\hline
F0-F6				&	658	&	22.16	&	15.92	&	12.11	&	29.85 \\
G5-G8 ($t > 8$ Gyr)	&	143	&	29.20	&	21.19	&	18.09	&	40.36 \\
\hline
\end{tabular}
\end{table*}

\begin{figure*}[h]
\begin{center}
\includegraphics[scale=0.3, angle=0]{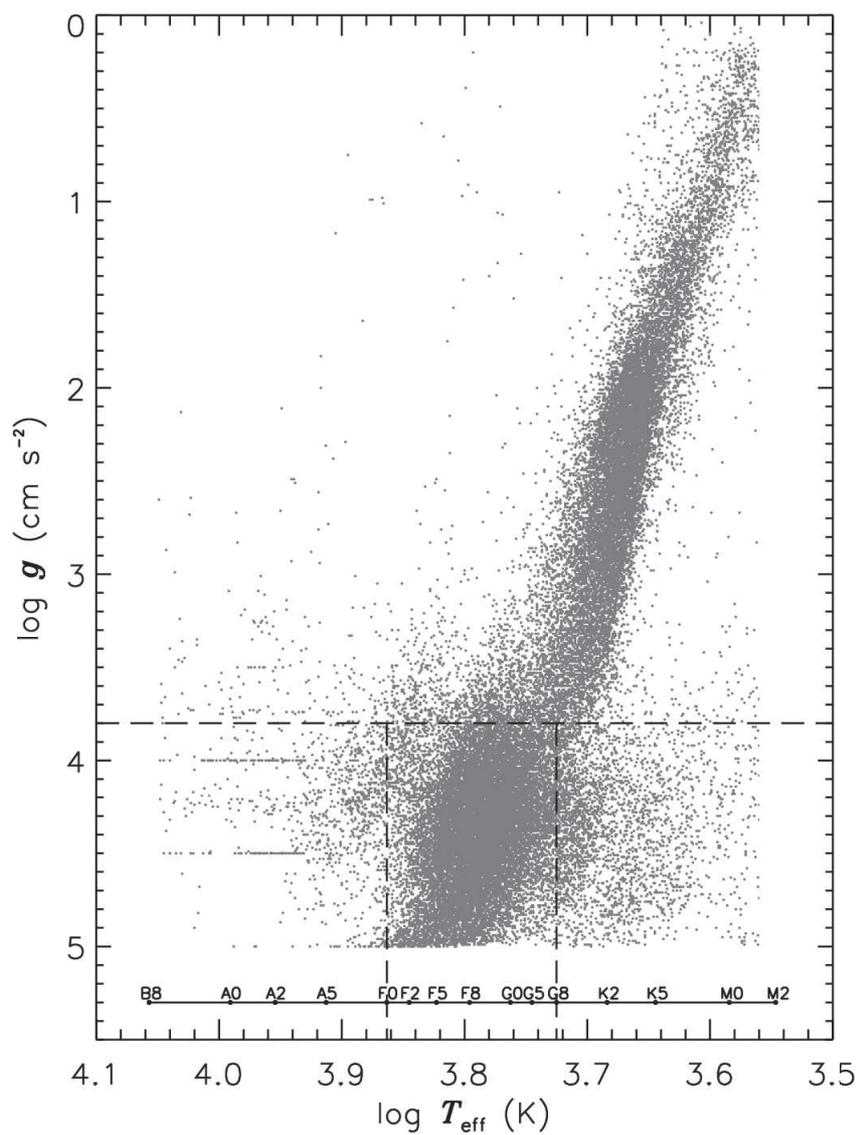}
\caption[] {Distribution of (all) DR3 stars in the $\log g$, $\log T_{eff}$ 
plane. Spectral types are also shown in the horizontal axis.}
\end{center}
\end{figure*} 

\begin{figure*}[h]
\begin{center}
\includegraphics[scale=.3, angle=0]{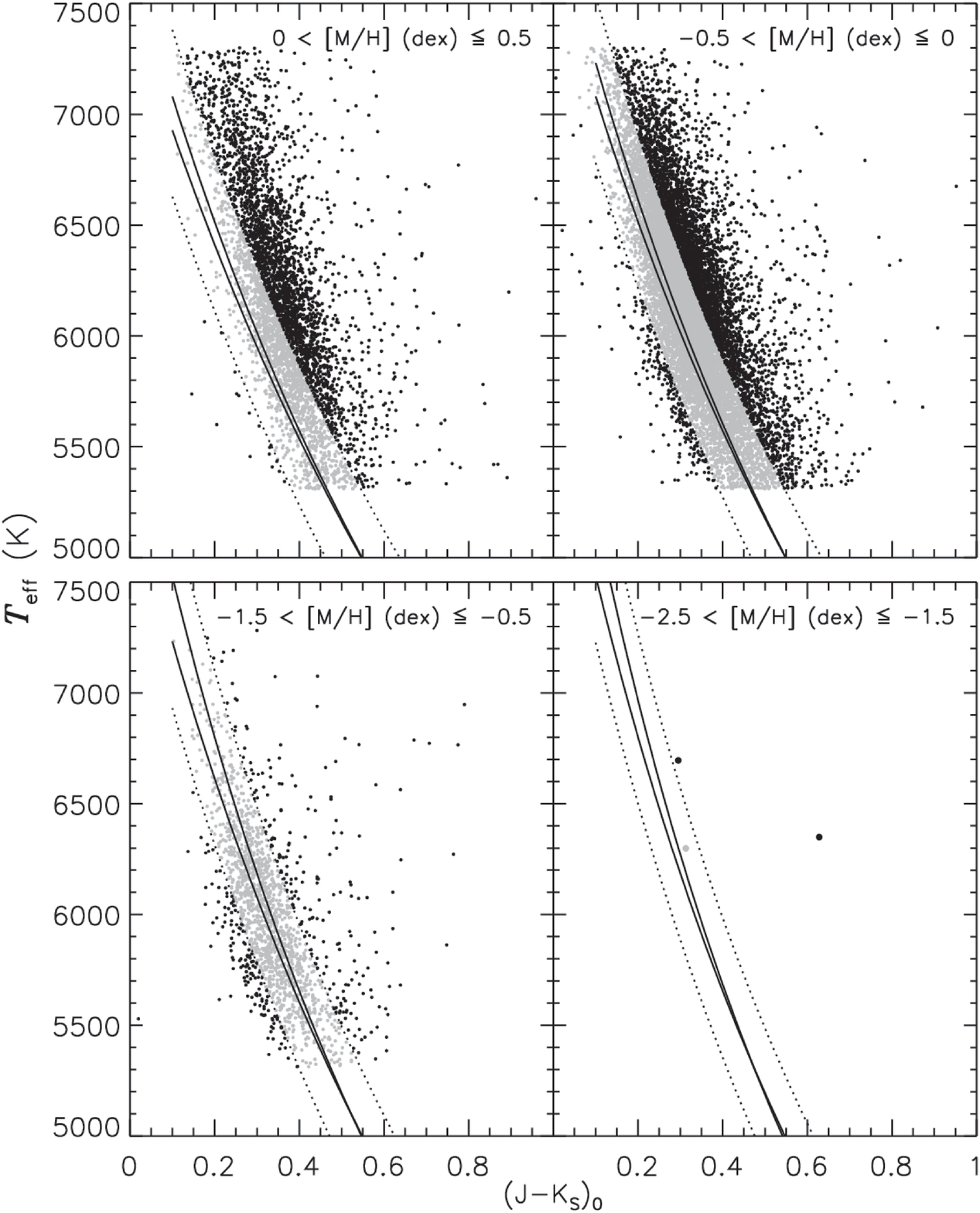}
\caption[] {The $T_{eff}$-$(J-K_s)_0$ diagram of F and G main sequence stars 
in four metallicity intervals. The rigid lines show the region occupied by 
the stars of \citet{Gonzalez09}, while the dotted ones indicate the 
2$\sigma$ dispersion of the mean metallicity in each panel. The rigid 
line on the left indicates the lower metallicity limit in the corresponding 
panel while the right one denotes the higher metallicity limit in the same 
panel. The two upper panels show that there are differences between the 
metallicities evaluated in the RAVE DR3 and in \citet{Gonzalez09}.}
\end{center}
\end{figure*} 

\begin{figure*}[h]
\begin{center}
\includegraphics[scale=.3, angle=0]{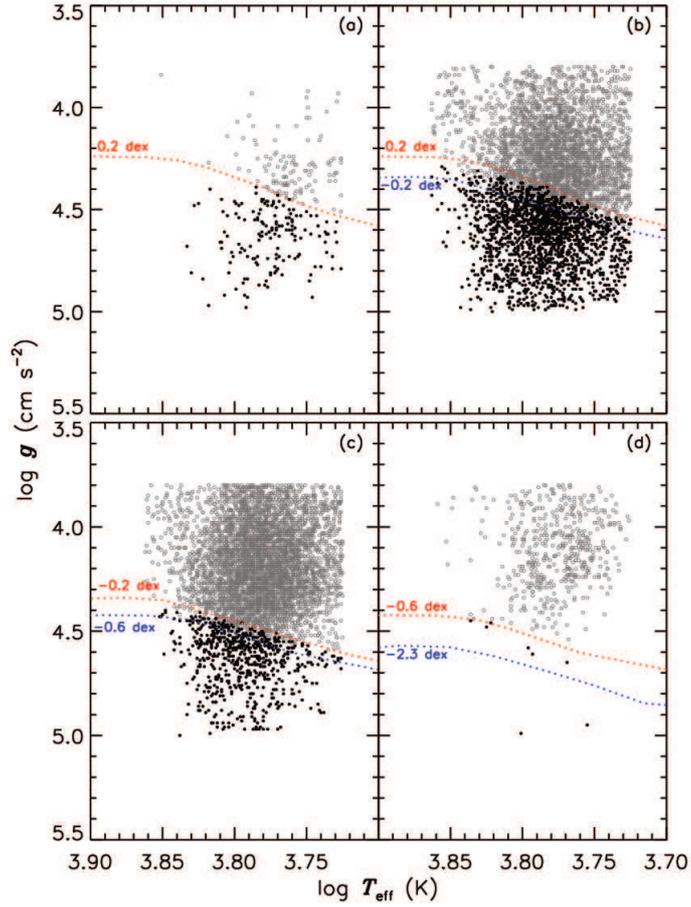}
\caption[] {Position of the star sample in four metallicity intervals, 
$0.2\le[M/H]$, $-0.2\leq[M/H]<0.2$, $-0.6\leq[M/H]<-0.2$, $[M/H]<-0.6$ dex, 
relative to the ZAMS Padova isochrone. Stars which fall below the ZAMS were 
omitted. The large fraction of stars below the ZAMS are due to the unexpected 
large values of surface gravity. However, exclusion of these stars do not 
affect our results. Because, this scattering is valid for all metallicity 
intervals.}
\end{center}
\end{figure*} 

\begin{figure*}[h]
\begin{center}
\includegraphics[scale=.26, angle=0]{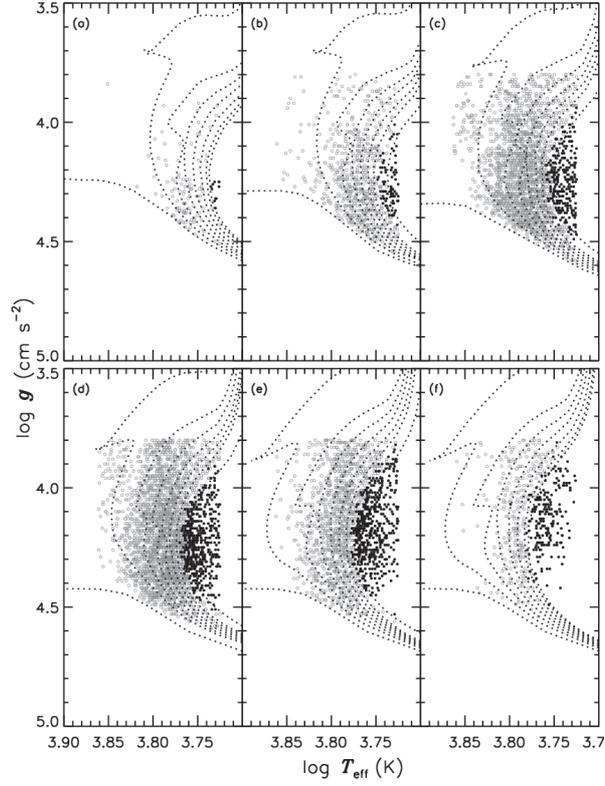}
\caption[] {Star sample in six metallicity intervals,
(a) $0.2\le[M/H]$, (b) $0\le[M/H]<0.2$, (c) $-0.2\le[M/H]<0$, (d)
$-0.4\le[M/H]<-0.2$, (e)$-0.6\le[M/H]<-0.4$, (f) $[M/H]<-0.6$ dex, 
fitted to Padova isochrones.}
\end{center}
\end{figure*} 

\begin{figure*}[h]
\begin{center}
\includegraphics[scale=0.2, angle=0]{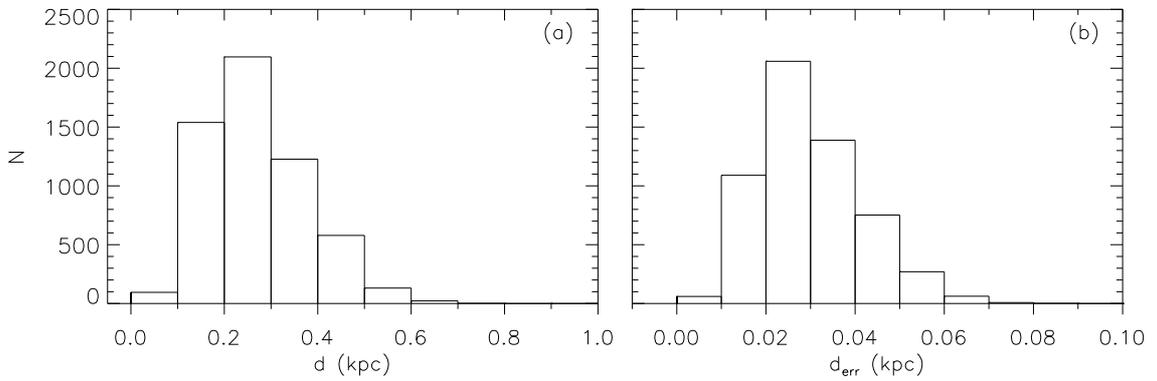}
\caption[] {Frequency (a) and error (b) distributions of distances
of F-G main sequence stars.}
\end{center}
\end{figure*} 

\begin{figure*}[h]
\begin{center}
\includegraphics[scale=0.25, angle=0]{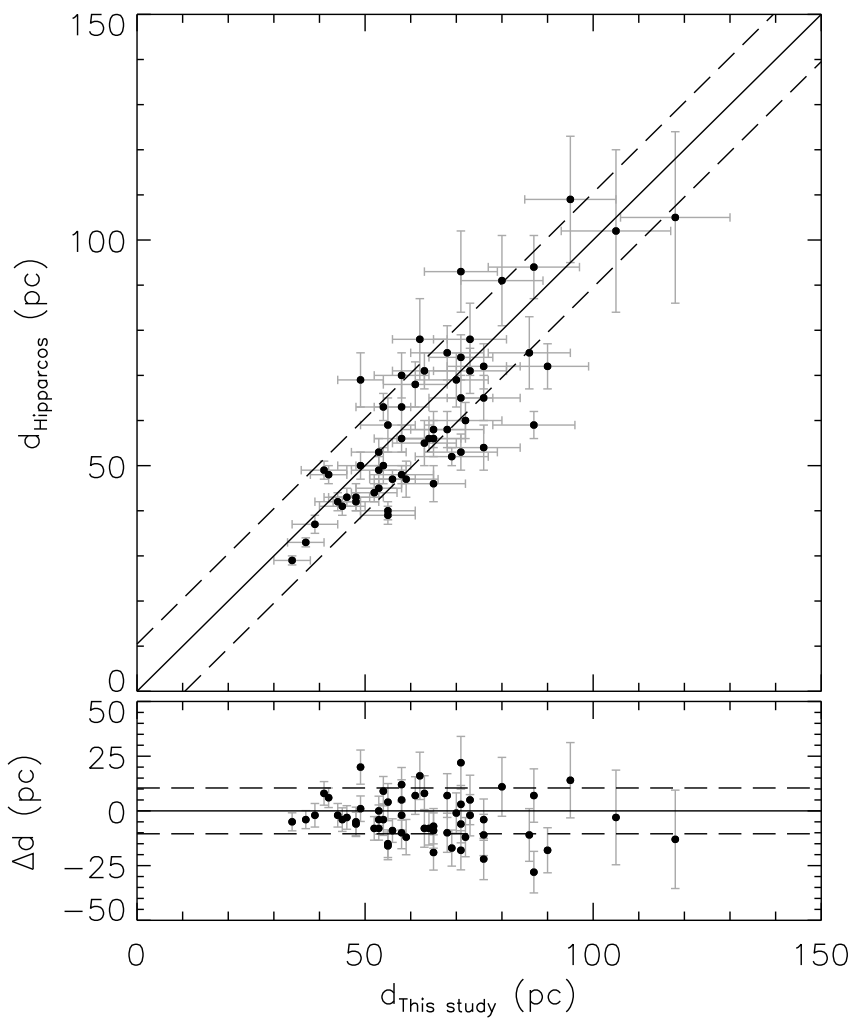}
\caption[] {Comparison of the distances estimated in our work with the ones 
evaluated by means of their parallaxes taken from {\em Hipparcos} catalogue. 
The one to one line is also given in the figure.}
\end{center}
\end{figure*} 

\begin{figure*}[h]
\begin{center}
\includegraphics[scale=0.15, angle=0]{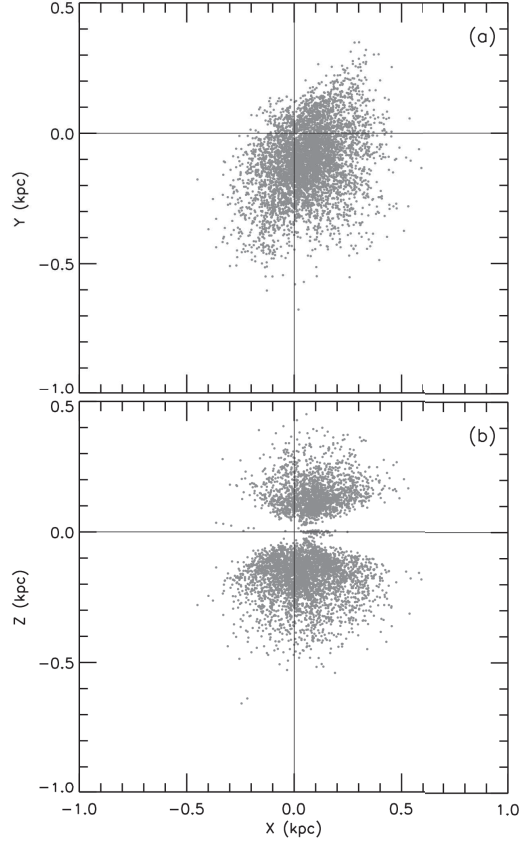}
\caption[] {Space distributions of RAVE DR3 F-G main sequence stars on
two planes. (a) $X-Y$ and (b) $X-Z$.}
\end{center}
\end{figure*}

\begin{figure*}[h]
\begin{center}
\includegraphics[scale=0.20, angle=0]{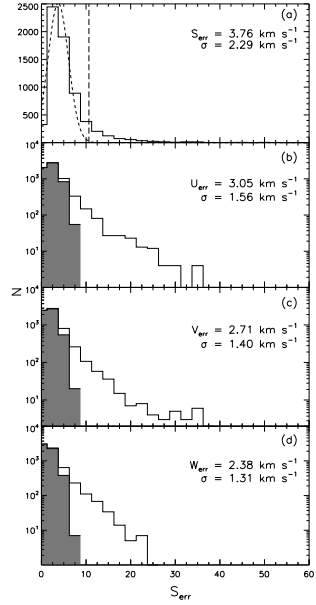}
\caption[] {Error histograms for space velocity (a) and its components
(b-d) for RAVE DR3 F-G main sequence stars. The vertical dashed line  
in panel (a) indicates the upper limit of the total error adopted in 
this work.The shaded part of the histogram indicates the error for 
different velocity components of stars after removing the stars with 
large space velocity errors.}
\end{center}
\end{figure*} 

\begin{figure*}[h]
\begin{center}
\includegraphics[scale=0.20, angle=0]{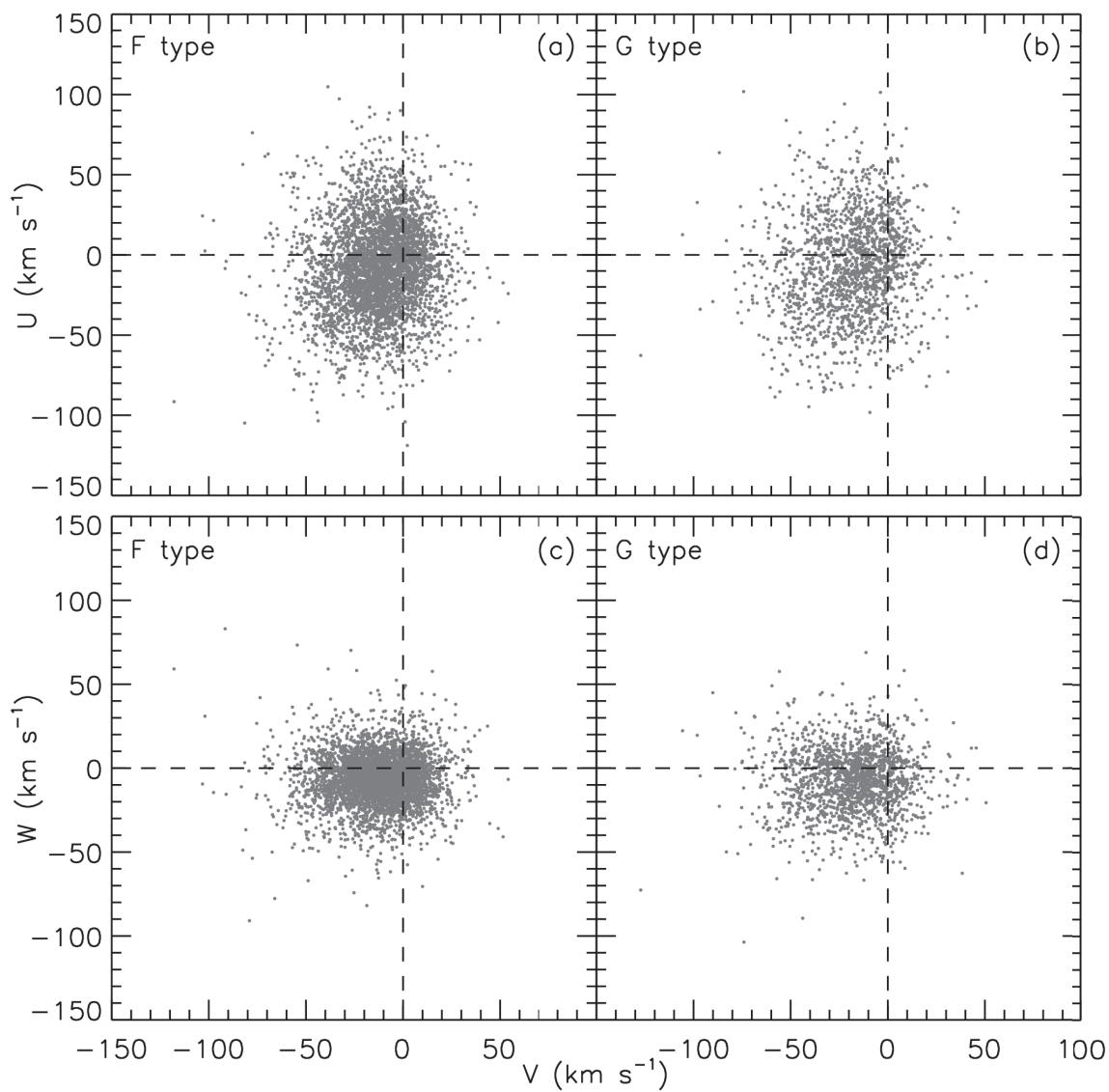}
\caption[] {The distribution of velocity components of our final sample 
of RAVE DR3 F-G main sequence stars with high-quality data, in two 
projections: $U-V$ (a and c) and $W-V$ (b and d).}
\end{center}
\end{figure*} 

\begin{figure*}[h]
\begin{center}
\includegraphics[scale=0.25, angle=0]{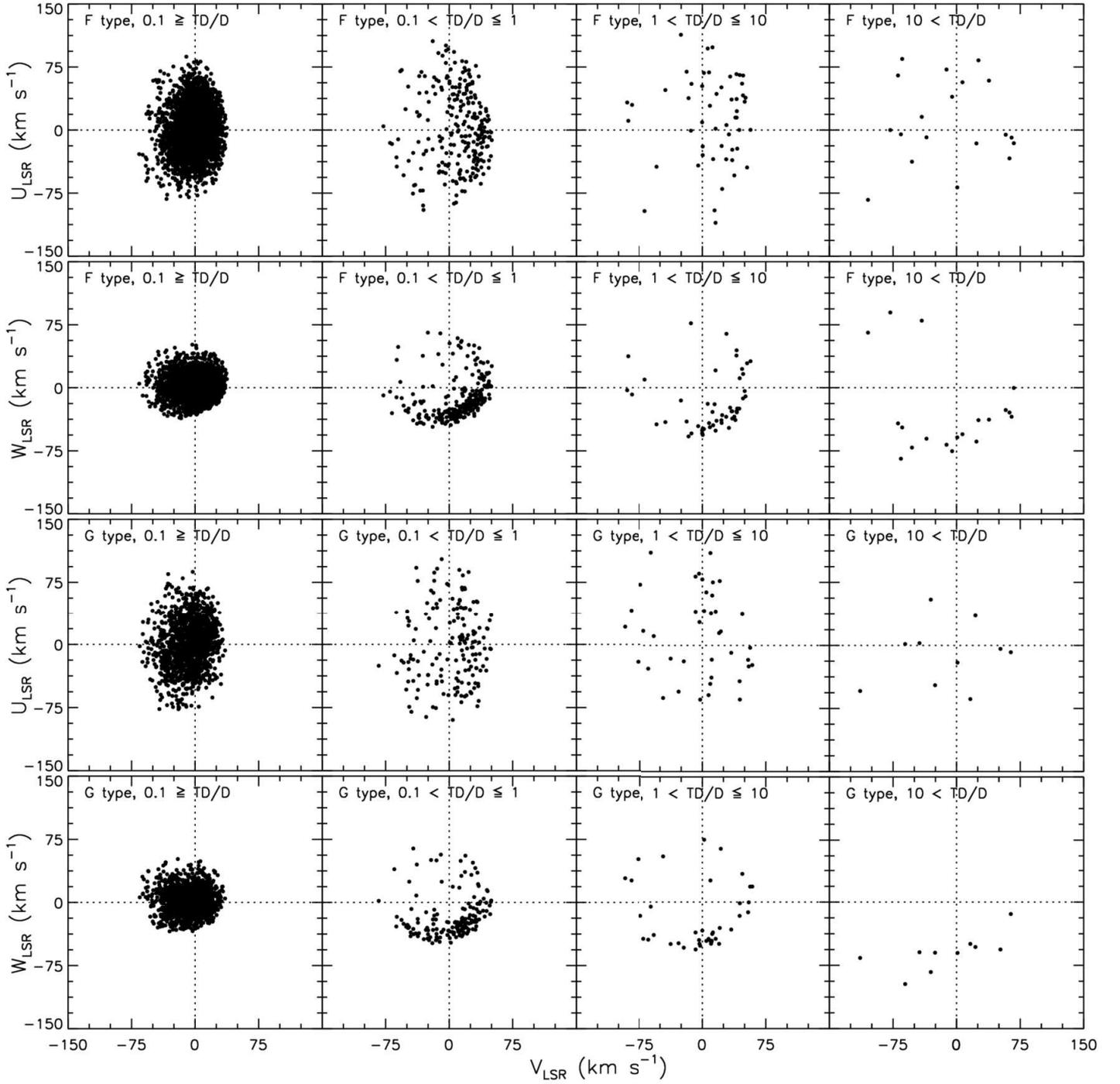}
\caption[] {$U-V$ and $W-V$ diagrams of F- and G-type stars applying 
Bensby et al.'s (2003) population classification criteria. It is seen 
that space-motion uncertainties remain significant, even for this 
nearby sample.}
\end{center}
\end{figure*}

\begin{figure*}[h]
\begin{center}
\includegraphics[scale=0.15, angle=0]{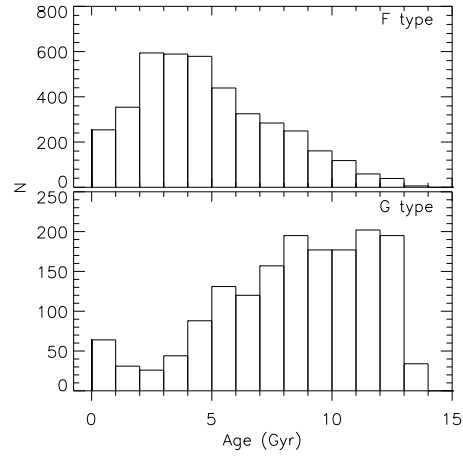}
\caption[] {Age distribution of the RAVE DR3 F and G main sequence stars.}
\end{center}
\end{figure*}

\begin{figure*}[h]
\begin{center}
\includegraphics[scale=0.15, angle=0]{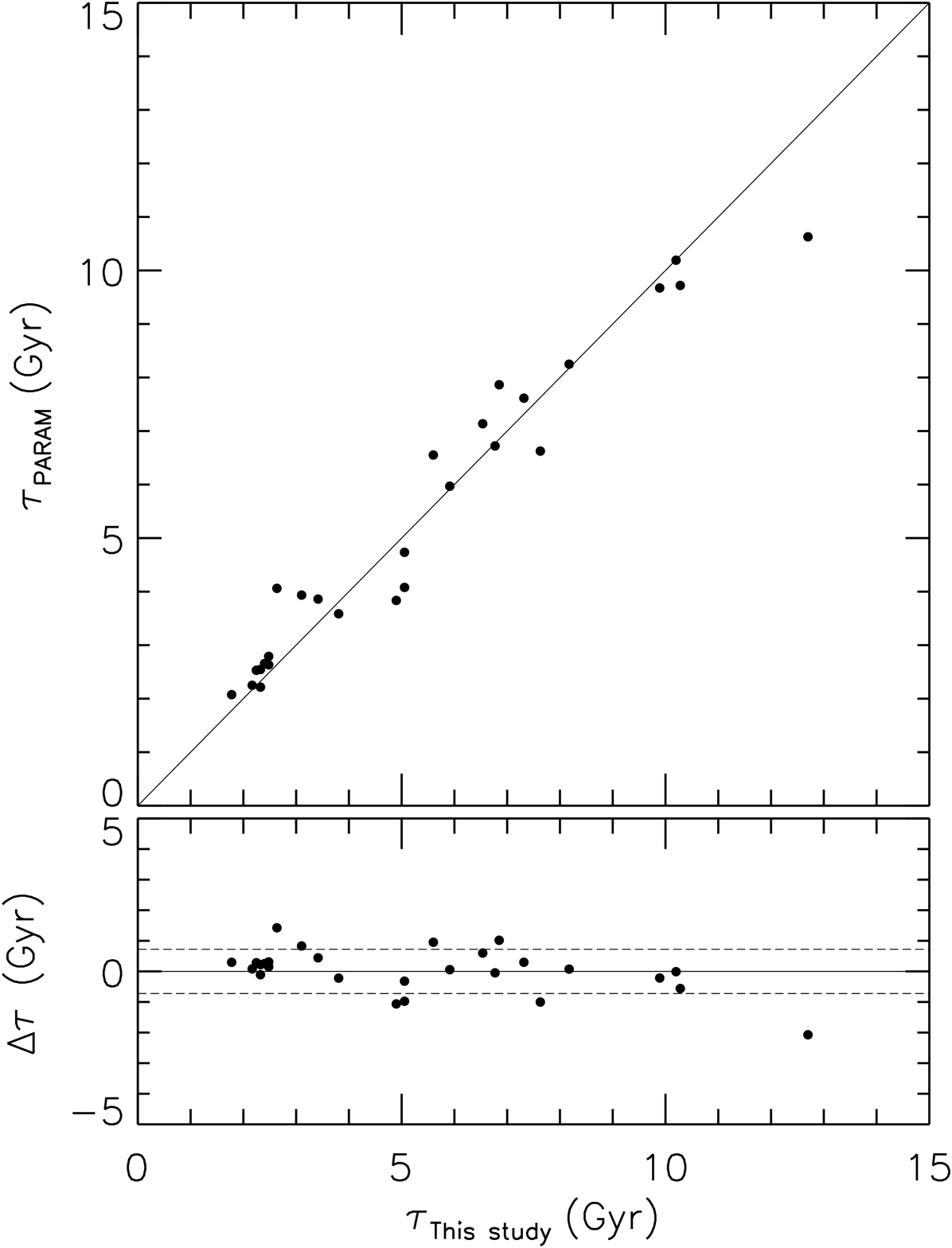}
\caption[] {Comparison of the ages for 25 stars estimated in our study 
with the ones evaluated by applying the PARAM webpage to the data $T_{eff}$, 
$M_V$ and $[M/H]$ taken from \cite{Holmberg09}. The dashed lines indicate 
$\pm$ 1$\sigma$ limits.}
\end{center}
\end{figure*}

\begin{figure*}[h]
\begin{center}
\includegraphics[scale=0.35, angle=0]{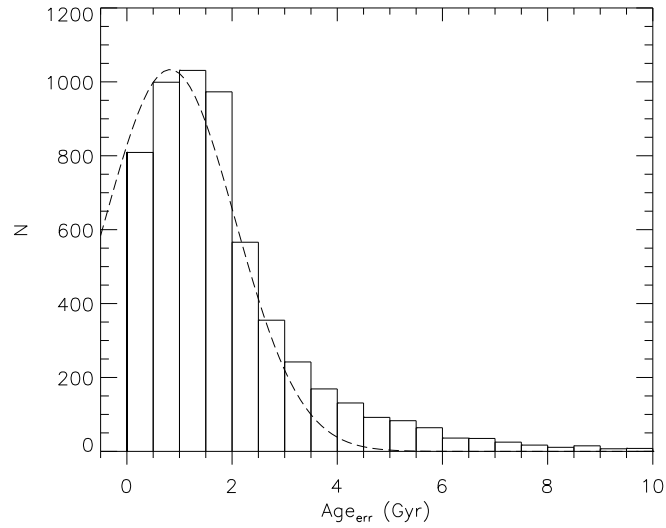}
\caption[] {Distribution of errors for ages estimated in our study. 
The dashed curve indicates the Gaussian distribution.}
\end{center}
\end{figure*}

\begin{figure*}[h]
\begin{center}
\includegraphics[scale=0.25, angle=0]{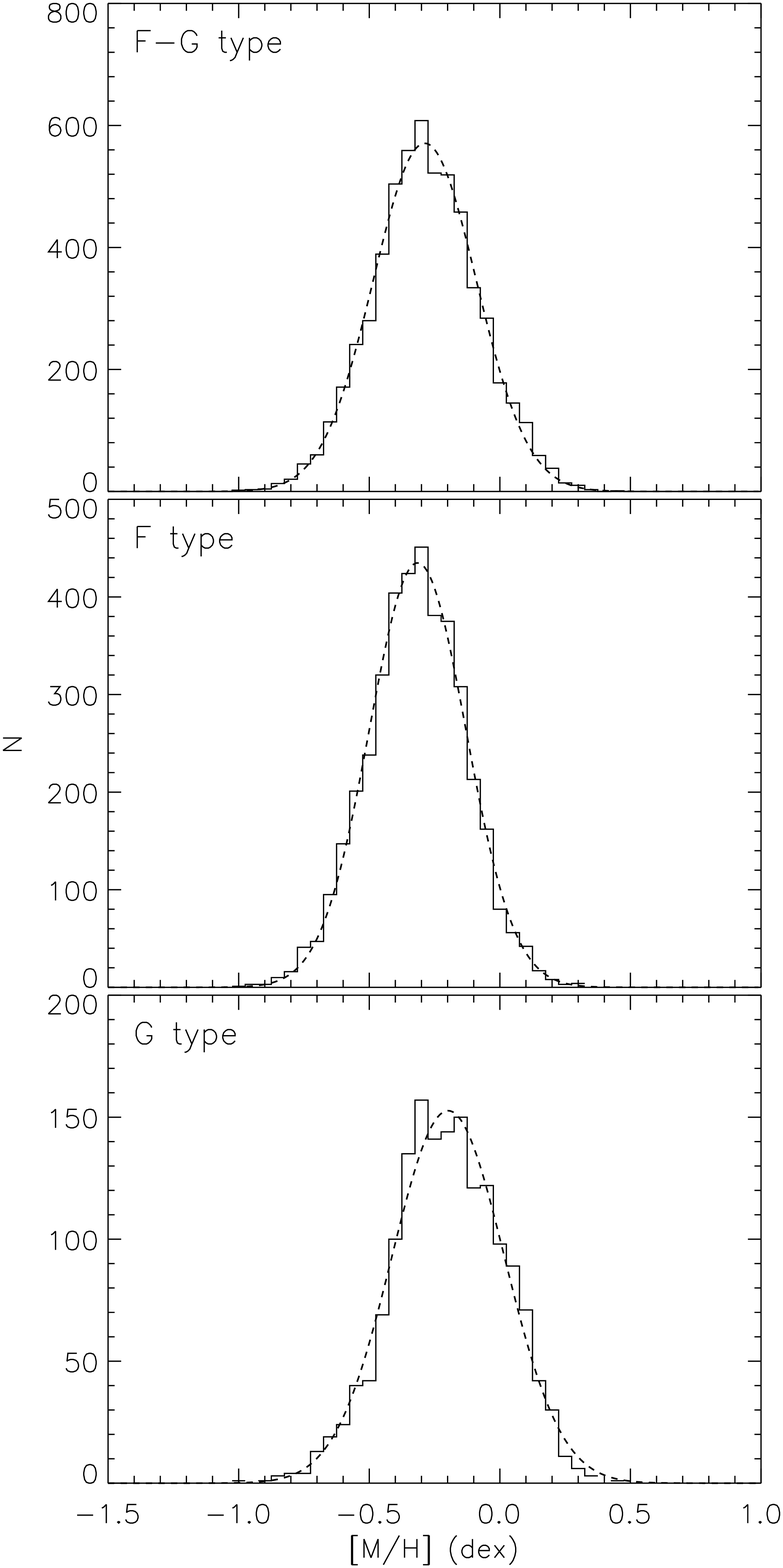}
\caption[] {Distribution of metallicities of the star sample as 
a function of spectral type, fitted to a Gaussian distribution.}
\end{center}
\end{figure*}

\begin{figure*}[h]
\begin{center}
\includegraphics[scale=.20, angle=0]{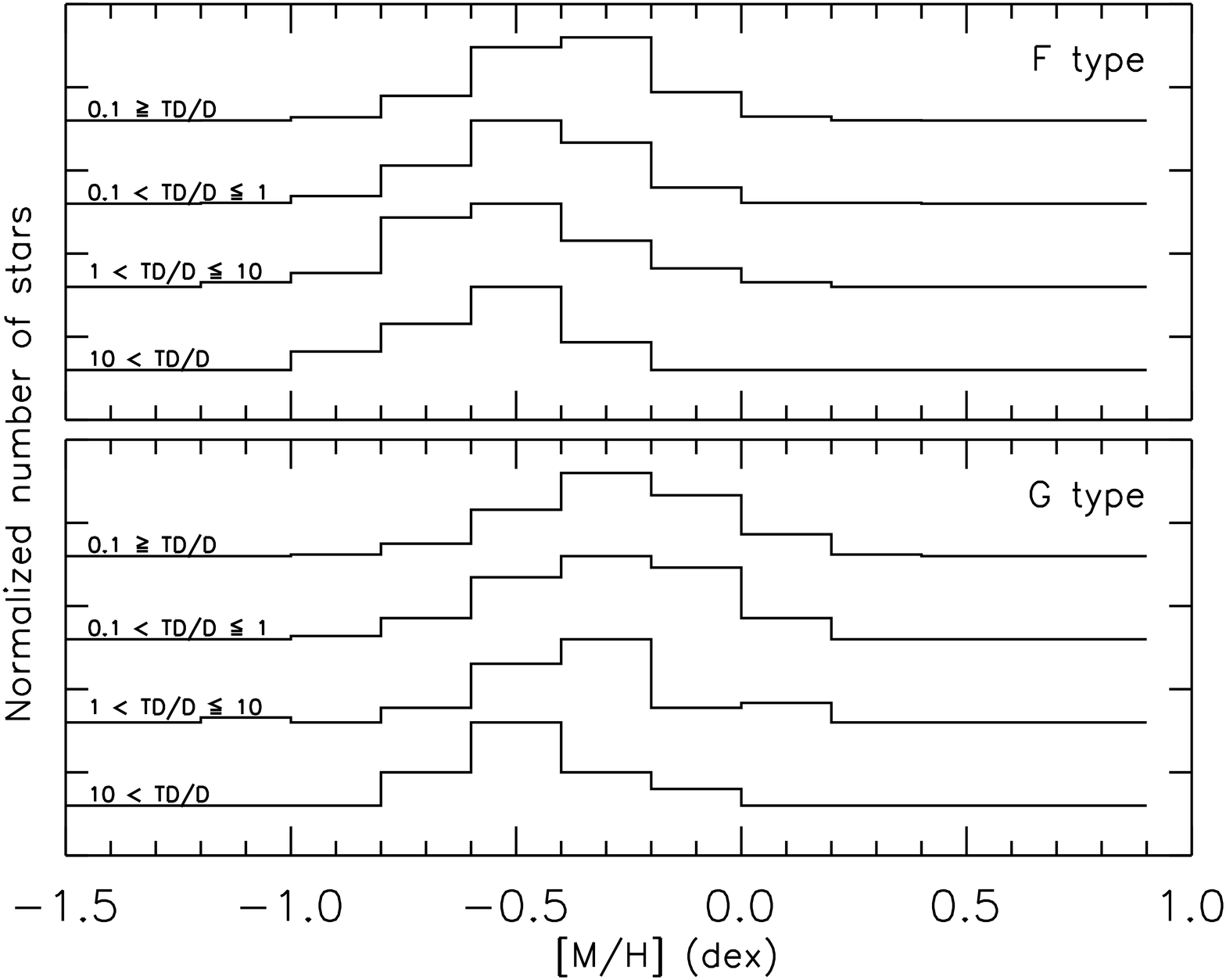}
\caption[] {Normalized metallicity distributions for different 
populations of F and G type main sequence stars.}
\end{center}
\end{figure*}

\begin{figure*}[h]
\begin{center}
\includegraphics[scale=.2, angle=0]{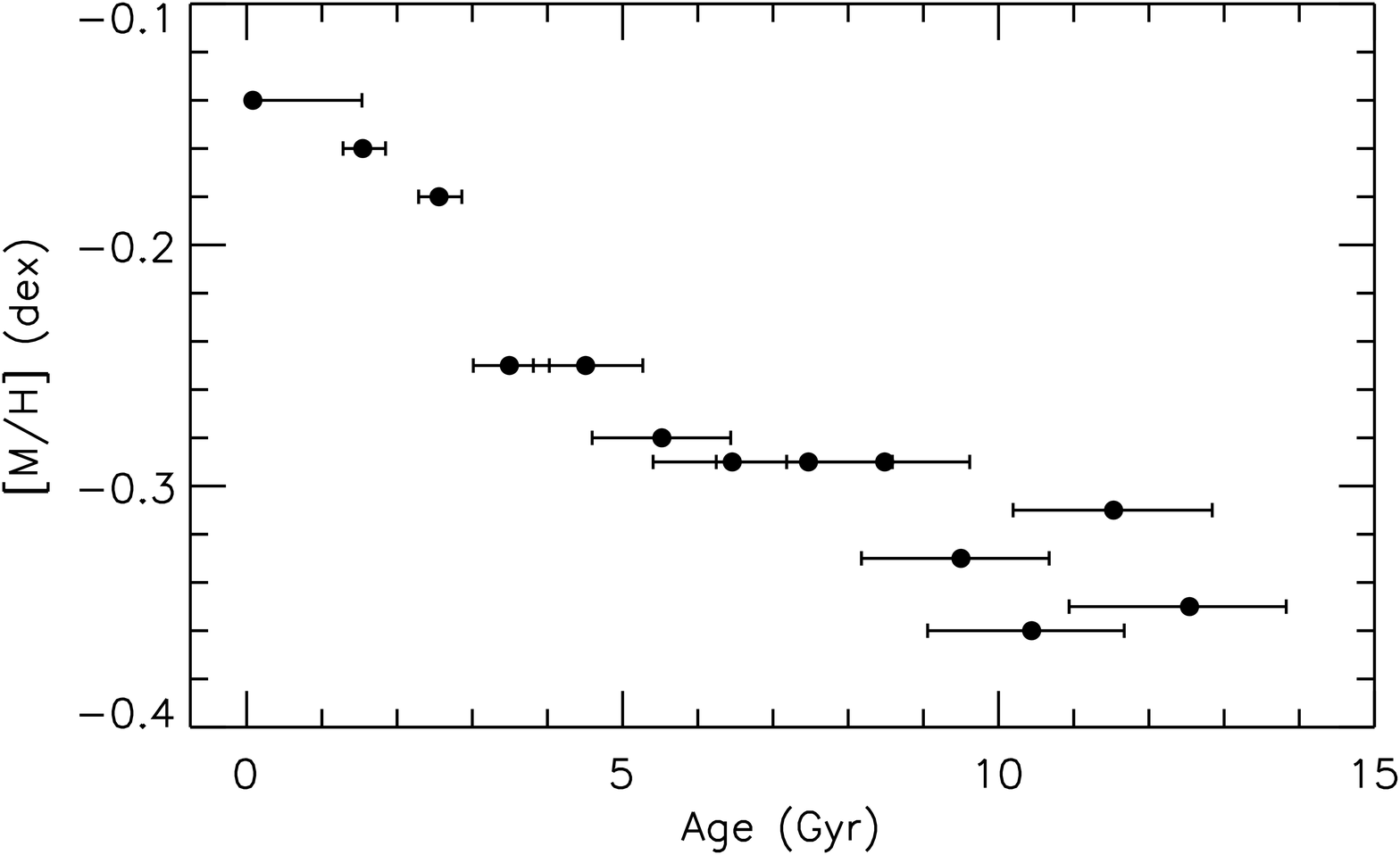}
\caption[] {Age-metallicity distribution of (all) sample stars.}
\end{center}
\end{figure*}

\begin{figure*}[h]
\begin{center}
\includegraphics[scale=.25, angle=0]{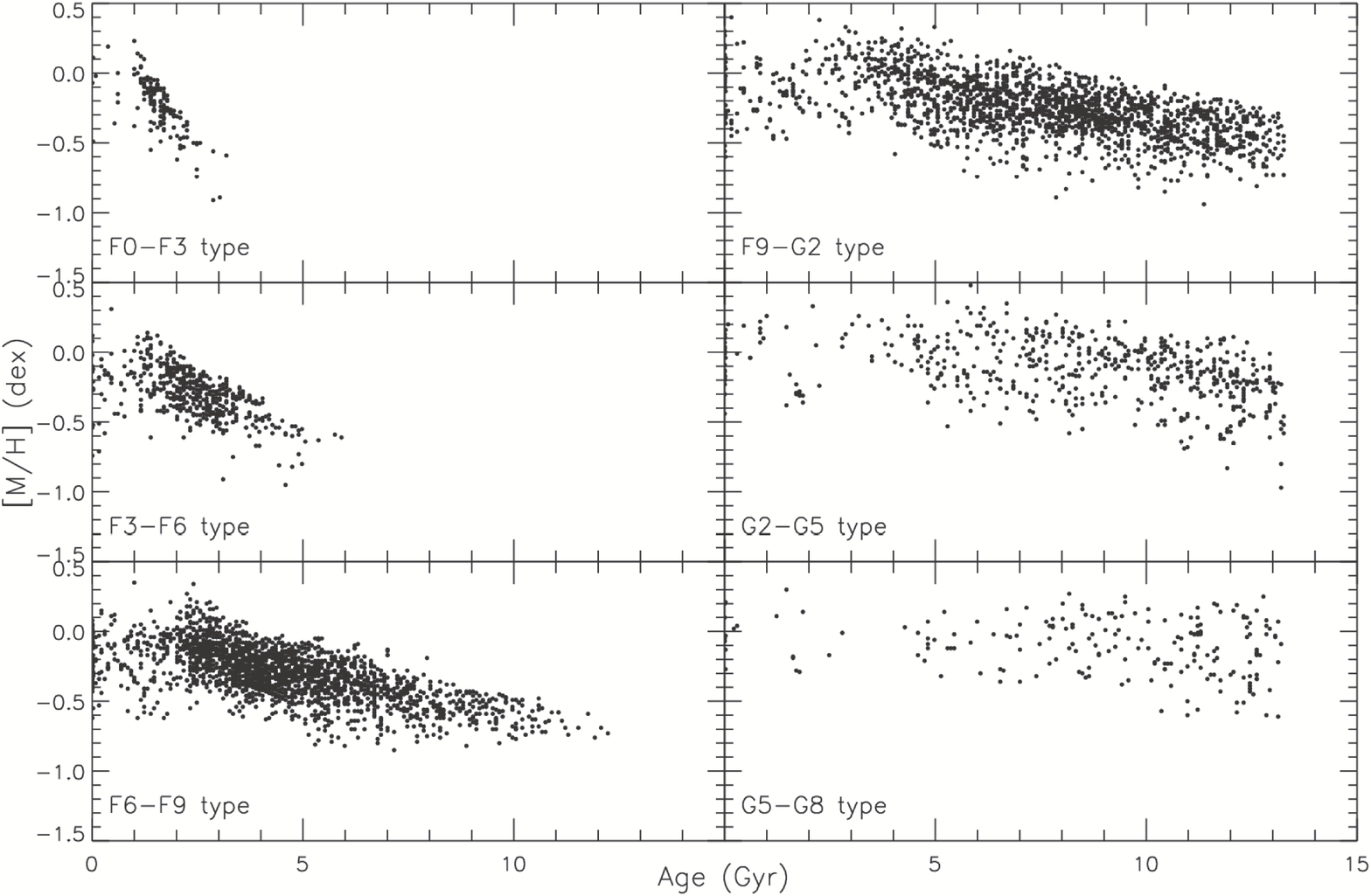}
\caption[] {Age-metallicity relation as a function of spectral type 
as indicated in six panels.}
\end{center}
\end{figure*}

\begin{figure*}[h]
\begin{center}
\includegraphics[scale=.25, angle=0]{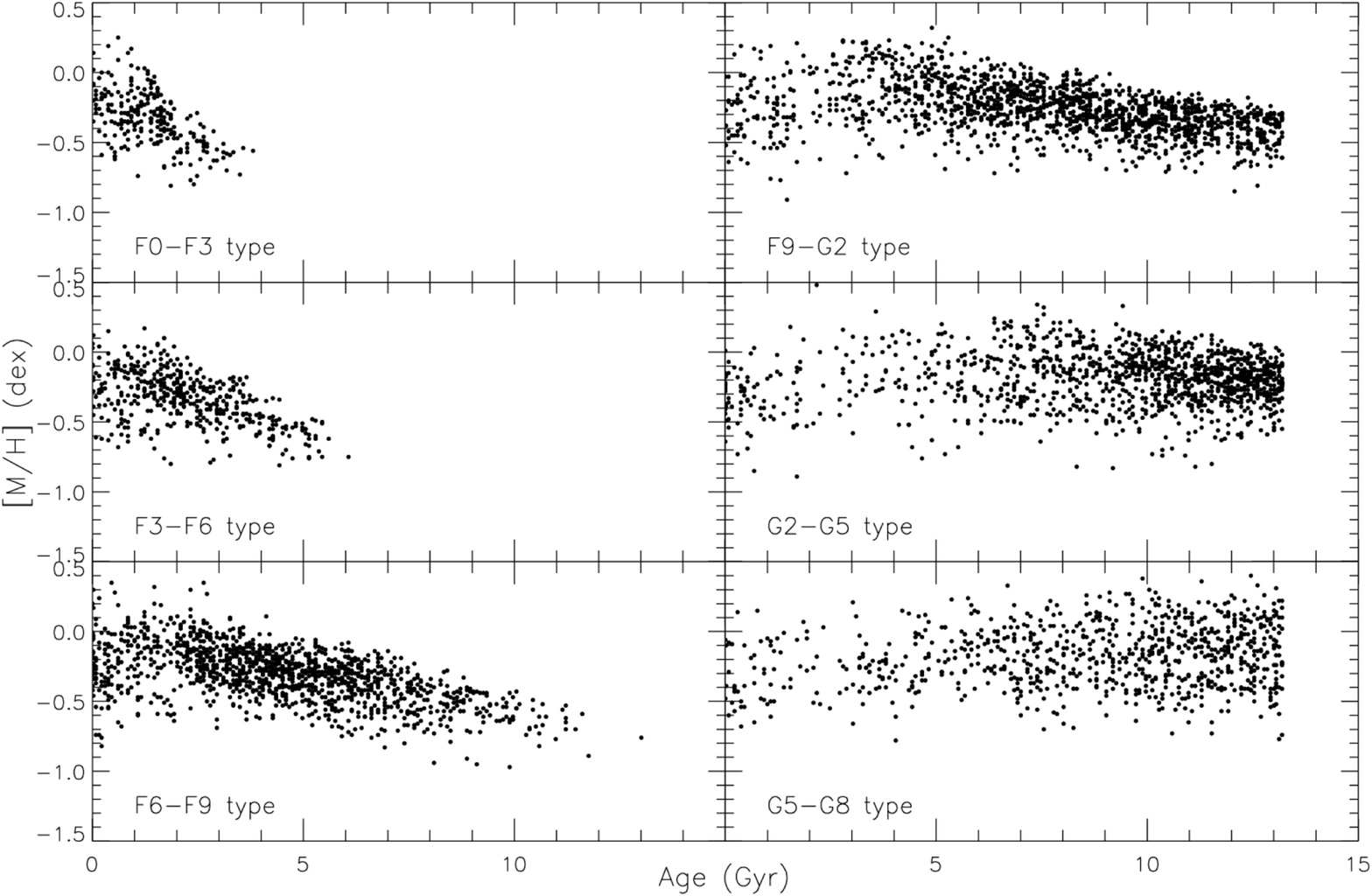}
\caption[] {``Artificial'' age-metallicity relation as a function of 
spectral type as indicated in six panels.}
\end{center}
\end{figure*}

\begin{figure*}[h]
\begin{center}
\includegraphics[scale=0.15, angle=0]{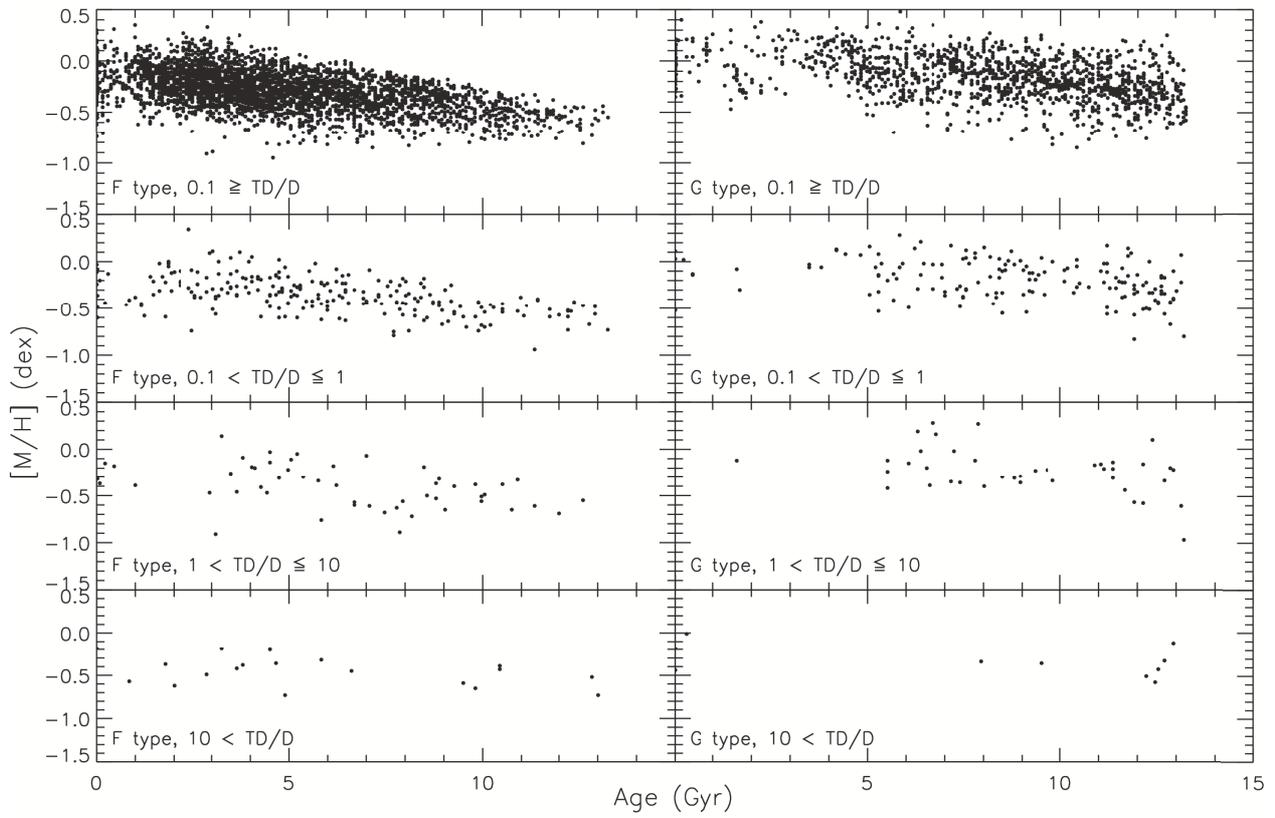}
\caption[] {Age-metallicity distribution as a function of population for 
F and G type stars as indicated in eight panels.}
\end{center}
\end{figure*}

\begin{figure*}[h]
\begin{center}
\includegraphics[scale=0.25, angle=0]{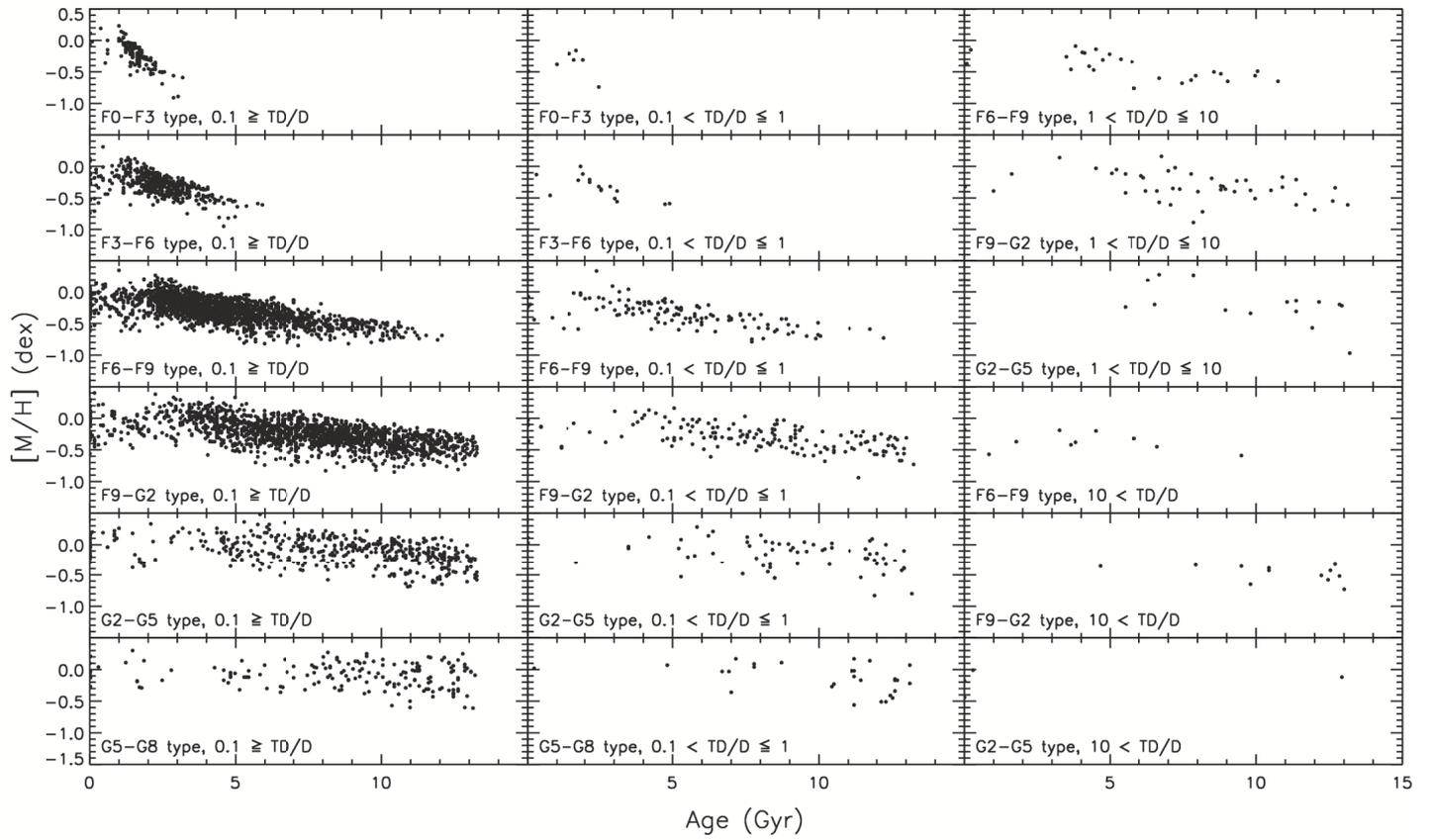}
\caption[] {Age-metallicity relations as a function of both spectral type 
and population as indicated in 18 panels.}
\end{center}
\end{figure*}

\begin{figure*}[h]
\begin{center}
\includegraphics[scale=0.25, angle=0]{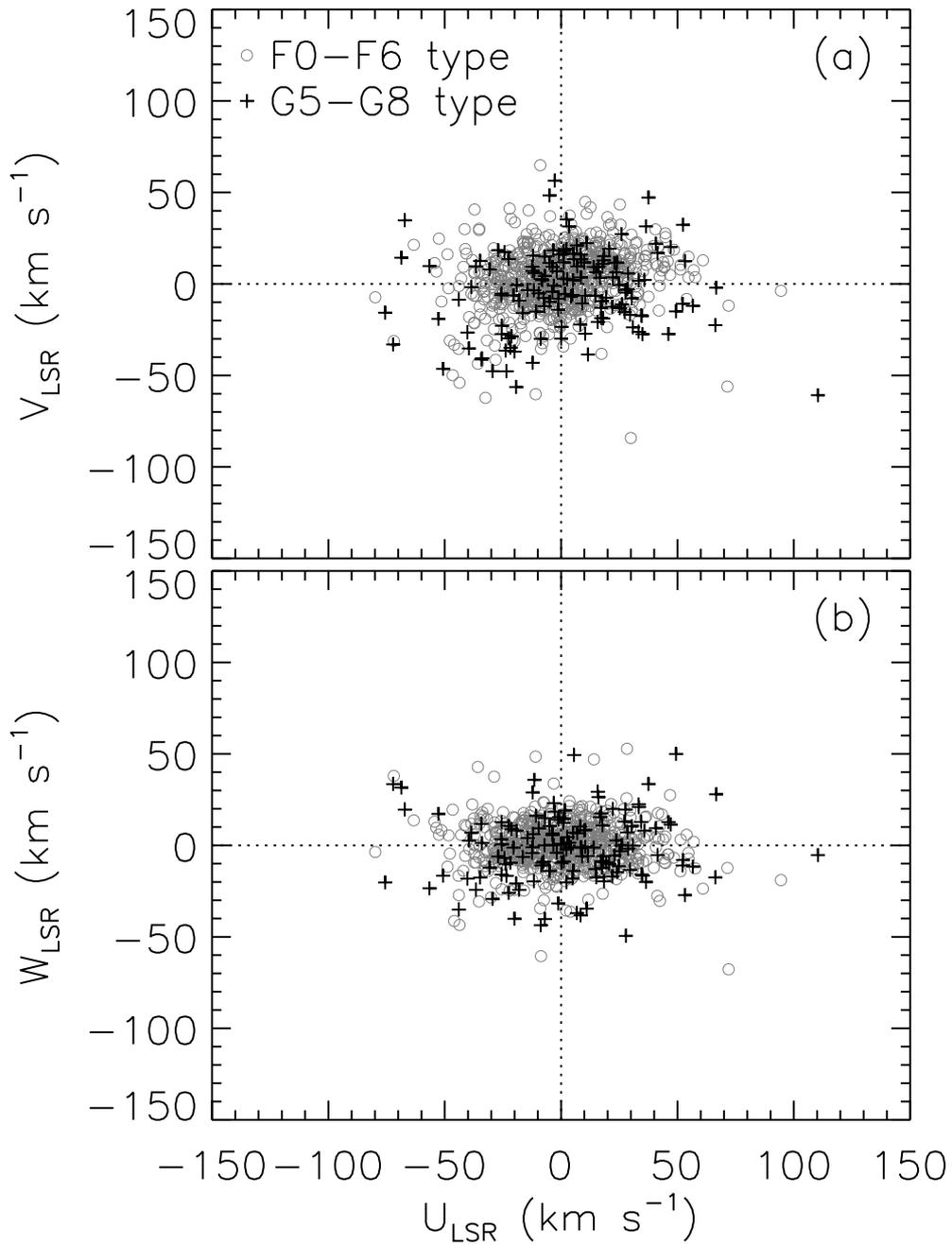}
\caption[] {Distribution of F0-F6 ($\circ$) and G5-G8 (+) spectral type stars in the  space velocity component planes in two panels: (a) ($U$, $V$) and (b) ($U$, $W$).}
\end{center}
\end{figure*}

\end{document}